\documentclass[epj,final]{svjour}
\usepackage{graphicx}

\newcommand{\beq}{\begin{equation}}
\newcommand{\beqa}{\begin{eqnarray}}
\newcommand{\eeq}{\end{equation}}
\newcommand{\eeqa}{\end{eqnarray}}

\newcommand{\abs}[1]{\vert#1\vert}
\newcommand{\bin}[2]{{#1\choose#2}}
\newcommand{\dy}{{\rm d}}
\newcommand{\e}{{\rm e}}
\newcommand{\eps}{\varepsilon}
\newcommand{\exi}{x_{\rm int}}
\newcommand{\f}{{\rm f}}
\renewcommand{\frac}[2]{\displaystyle{\displaystyle#1\over\displaystyle#2}}
\newcommand{\gs}{{\rm gs}}
\newcommand{\lra}{\Longrightarrow}
\newcommand{\mean}[1]{\langle#1\rangle}

\renewcommand{\o}{\omega}
\newcommand{\s}{\sigma}
\newcommand{\st}{{\rm s}}
\newcommand{\sign}{\mathop{\rm sign}\nolimits}
\newcommand{\var}{\mathop{\rm var}\nolimits}
\newcommand{\xidy}{\xi_{\rm dyn}}
\newcommand{\xii}{\xi_{\rm int}}
\newcommand{\xiic}{\xi_{{\rm int,}c}}
\newcommand{\C}{{\cal C}}
\newcommand{\E}{{\cal E}}
\newcommand{\Int}{\mathop{\rm Int}\nolimits}
\renewcommand{\L}{{\cal L}}
\newcommand{\N}{{\cal N}}
\renewcommand{\S}{\Sigma}
\newcommand{\T}{\Gamma}

\begin{document}

\title{The effects of grain shape and frustration in a granular column near
jamming}

\author{J.M.~Luck\inst{1} \and A.~Mehta\inst{2}}

\institute{Institut de Physique Th\'eorique, IPhT, CEA Saclay,
and URA 2306, CNRS, 91191 Gif-sur-Yvette cedex, France.\\
\email{jean-marc.luck@cea.fr}
\and
S.N.~Bose National Centre for Basic Sciences, Block JD, Sector 3,
Salt Lake, Calcutta 700098, India.\\
\email{anita@bose.res.in}}

\date{}

\abstract
{We investigate the full phase diagram of a column of grains near jamming,
as a function of varying levels of frustration.
Frustration is modelled by the effect of two opposing fields on a grain,
due respectively to grains above and below it.
The resulting four dynamical regimes
(ballistic, logarithmic, activated and glassy)
are characterised
by means of the jamming time of zero-temperature dynamics,
and of the statistics of attractors reached by the latter.
Shape effects are most pronounced in the cases of strong and weak
frustration, and essentially disappear around a mean-field point.}

\PACS{
{64.60.My}{Metastable phases}
\and
{45.70.Cc}{Static sandpiles; granular compaction}
\and
{45.70.Vn}{Granular models of complex systems; traffic flow}
\and
{64.70.Q-}{Theory and modeling of the glass transition}
}

\maketitle

\section{Introduction}
\setcounter{equation}{0}
\def\theequation{1.\arabic{equation}}

One of the reasons why heterogeneities are intrinsic to granular media
is the absence of thermal motion; spatial and temporal behaviour has no
reason to equilibrate, so that structures as well as time tracks which
are far out of equilibrium can remain juxtaposed in the same system.
Clearly, this leads to spatially and dynamically heterogeneous behaviour.
It is, however, only recently that research efforts in this context
have focused on heterogeneity (see~\cite{sm} for a recent review).
Examples of static spatial heterogeneity in granular systems
include bridges~\cite{br,pak,bridges} and force chains~\cite{science95}.
More generally, dynamical heterogeneities have attracted a lot of attention
recently,
both in granular matter (see~\cite{dhg})
and in other systems such as glasses or colloids (see~\cite{dhs}).

Spatiotemporal heterogeneity takes place when different
parts of a system have diverse dynamical behaviour characteristic of their
location.
The first indications of such heterogeneity in a vibrated granular medium
were found in the experiments of Reference~\cite{sid}, the findings of which
indicated that
both the average density, as well as density fluctuations, varied strongly
throughout a shaken box of grains as a function of depth.
Computer simulation and theoretical results~\cite{VI} reproduced this
behaviour,
predicting additionally that the mean density was an increasing function of
depth, and that density fluctuations were largest in the middle of the box for
time windows relevant to experiment.
These results additionally suggested that the phase behaviour within the box
was very heterogeneous; ballistic behaviour was expected near the top,
activated in the middle and glassy behaviour at its base.

The theoretical model~\cite{II,III,IV,V} on the basis of which
the latter predictions were made is as follows:
grains in a column are able to orient themselves in one of two possible ways,
corresponding to `ordered' and `disordered'.
When a grain is in its disordered orientation, space is wasted:
a void of size $\eps$ is created, which characterises the shape of the grain.
Rational and irrational values of $\eps$
correspond to regular and irregular grain shapes respectively.
Despite the simplicity of this `aspect ratio' formulation of shape
effects, recent work~\cite{del} has shown that it may be used
to characterise a rich variety of granular shapes.

The presence of gravity is included in the model by a depth-dependent local
frequency, such that lower (more weight-bearing) grains move more slowly
than the less burdened upper grains.
In the jamming limit, voids are minimized;
accordingly, the definition of a ground state in the model
is one that locally minimizes the voids ratio~\cite{br}.
However, since the
orientation that minimizes voids with respect to grains above a given grain
is not typically that which fulfils the same function for those below itself,
this naturally generates {\it frustration}.
In the model, this is represented by the effect of two oppositely directed
fields whose relative strengths are modulated by a coupling constant~$g$.

In earlier work~\cite{II,III,IV,V}, the focus was on the effect of shape;
the behaviour of the model for typical rational and irrational values of
the shape parameter $\eps$ was explored for $g=0$~\cite{III,IV}
and then in the $g\to0$ limit~\cite{V}.
Here we complete the analysis by looking at the effect
of varying the coupling constant $g$.
This is equivalent to varying the frustration,
and as will be shown, has wide-ranging effects on the behaviour of the model.

The plan of the paper is as follows.
The definition of the model is recalled in Section~2.
Section~3 is devoted to the statics of the model,
i.e., the number and the nature of its ground states.
We then address the properties of zero-temperature dynamics.
Section~4 contains an analysis of the dynamical phase diagram
and of the behaviour of the jamming time in the various regimes,
whereas the statistics of attractors is investigated in Section~5.
In Section~6 we conclude with a brief discussion of our findings.

\section{The model}
\setcounter{equation}{0}
\def\theequation{2.\arabic{equation}}

In its most complete form, the model~\cite{V} consists of a finite column of
$N$ grains, labeled by their depth $n=1,\dots,N$.
Each grain has an orientation variable $\s_n=\pm1$.
Grain $n$ is referred to as {\it up} or {\it ordered} when $\s_n=+1$
and {\it down} or {\it disordered} when $\s_n=-1$.
Disordered orientations generate voids and waste space,
whereas ordered ones do not.
Implicit in this description is the effect of shape,
which is most easily understood
in terms of the rectangular grains of aspect ratio $a$ considered
in~\cite{II}.
Grains aligned along their long edges (length 1)
result in a fully packed column,
whereas those perched on their short edges (length $a<1$)
leave voids of size $\eps=1-a$.
The horizontal orientation is thus ordered, and the vertical one disordered.
Such a two-state model is clearly an approximation; we lump the effects of
all possible void spaces created by disordered orientations of arbitrarily
shaped grains into one disordered (vertical) orientation, and make
a similar approximation for the ordered (horizontal) orientation.

The $N$ binary variables $\{\s_n=\pm1\}$
define the $2^N$ configurations of the system.
We consider the following continuous-time stochastic dynamics
which do not obey detailed balance.
Grain orientations are updated with the Markovian rates
\beq
\left\{\matrix{
w(\s_n=+1\to\s_n=-1)=\e^{-(\lambda_n+H_n)/\T},\hfill\cr
w(\s_n=-1\to\s_n=+1)=\e^{-(\lambda_n-H_n)/\T},\hfill
}\right.
\label{rates}
\eeq
where

\smallskip\noindent$\bullet$
$\T$ is a dimensionless vibration intensity, referred to as temperature.

\smallskip\noindent$\bullet$
$\lambda_n$ is the activation energy of grain $n$,
which we take to be proportional to its depth:
\beq
\lambda_n=\frac{n\T}{\xidy}.
\eeq
The {\it dynamical length} $\xidy$
is the depth beyond which grains are frozen out by the sheer weight
of grains above them.
Thus, the frequency of response of a grain $n$
falls off exponentially with its depth:
\beq
\o_n=\e^{-\lambda_n/\T}=\e^{-n/\xidy}.
\label{odef}
\eeq

\smallskip\noindent$\bullet$
$H_n$ is the local ordering field felt by grain $n$,
which is determined by all the other grains, both above and below $n$.
The effect of the upper grains is assumed to be uniform.
The back-propagation from grains below a given grain cannot, of course,
be similarly uniform.
We assume for simplicity that upward constraints are exponentially damped,
with a characteristic length $\xii$, the {\it interaction length}.
We thus write
\beq
H_n=h_n+gj_n,
\label{hdef}
\eeq
where the uniform effect $h_n$ of grains above $n$ ($m=1$, $\dots$, $n-1$)
and the non-uniform effect $j_n$ of grains below~$n$ ($m=n+1,\dots,N$)
are given by
\beq
h_n=\sum_{m=1}^{n-1}f(\s_m),\quad
j_n=\sum_{m=n+1}^Nf(\s_m)\,\e^{-(m-n)/\xii},
\label{hjdef}
\eeq
whereas $g$ is a positive coupling constant.

Furthermore, both components $h_n$ and $j_n$
of the total local field $H_n$ acting on grain $n$
depend on every grain orientation $\s_m=\pm1$
through the same function, the `shape factor' $f(\s_m)$,
where
\beq
f(\s)=\frac{\eps-1}{2}-\frac{\eps+1}{2}\,\s=\left\{\matrix{
\eps\hfill&\hbox{if}\hfill&\s=-1,\cr
-1\;\hfill&\hbox{if}\hfill&\s=+1.}\right.
\label{fdef}
\eeq

The parameter $\eps$ can be thought of as the size of a typical void space
for a grain of a particular shape,
with rational and irrational values of $\eps$ corresponding to regular and
irregular grain shapes respectively.
There is an exact symmetry between models with $\eps$ and $1/\eps$,
so that the shape parameter can be restricted to $0\le\eps\le1$.

Putting all of this together, we see that
that the contribution of an ordered grain to the local field is (negative)
unity, while that of a disordered grain is a void space of magnitude $\eps$;
the latter is clearly
a function of granular shape, hence the name given to the shape factor
$f(\s_m)$.
The minus sign in the ordered case ensures that the contribution of an ordered
grain to the excess void space $H_n$ is {\it less} than that of a disordered
grain, as it ought to be; more importantly, this says that a void space is
destroyed every time a grain aligns in an ordered fashion relative to its
neighbours.

The dynamics of the model involves
the often conflicting contributions of the opposing local fields
$h_n$ and $j_n$.
In turn, these comprise all the terms $f(\s_m)$ which
take values $\eps$ or $-1$ according to~(\ref{fdef}), for all the
other grains~$m$ in the column.
As mentioned above, this represents a simple-minded way
of incorporating frustration into the model.

In the following, we use the notation
\beq
\exi=\e^{-1/\xii}.
\eeq
We mention the following recursion relations:
\beq
h_n=h_{n-1}+f(\s_{n-1}),\quad
j_n=\exi\left(f(\s_{n+1})+j_{n+1}\right),
\label{hjrel}
\eeq
with $h_1=j_N=0$,
which provide a fast algorithm to evaluate the local fields.

To sum up, the model parameters are the number of grains $N$,
the shape parameter $\eps$, the coupling constant~$g$,
and the interaction and dynamical lengths $\xii$ and $\xidy$.
Previous work has been devoted to investigations
of zero-temperature static (number and structure of ground states)
and dynamic (recovery of ground states as attractors) properties
of the model in several special cases of interest:
the directed model ($g=0$) for $\xidy=\infty$~\cite{III},
the directed model for general $\xidy$~\cite{IV},
and the weak-coupling regime ($g\ll1$) for $\eps=1$~\cite{V}.

In this work we aim at giving an overall picture of
zero-temperature properties of the model all over its parameter space,
with an emphasis on their dependence
on the coupling constant $g$ and on the lengths $\xii$ and $\xidy$.
As some of the features of the model are different for rational and irrational
values of the shape parameter $\eps$,
we shall use $\eps=1$ as our prototypical rational number,
and the (small) golden mean
\beq
\eps=\frac{1}{\Phi}=\frac{\sqrt{5}-1}{2}\approx0.618033
\label{gold}
\eeq
as our prototypical irrational number.

\section{Statics: ground states}
\setcounter{equation}{0}
\def\theequation{3.\arabic{equation}}

The rules~(\ref{rates}) simplify as follows
in the zero-temperature limit ($\T\to0$):
\beq
\frac{w(\s_n=-1\to\s_n=+1)}{w(\s_n=+1\to\s_n=-1)}=\e^{2H_n/\T}
\to\left\{\matrix{\infty\hfill&\hbox{if}\hfill&H_n>0,\hfill\cr
0\hfill&\hbox{if}\hfill&H_n<0.\hfill
}\right.
\label{zero}
\eeq
It is therefore natural to define a {\it ground state} as a configuration
where the orientation of every grain is aligned along its local
field~\cite{II,III,IV,V}:
\beq
\s_n=\sign H_n=\left\{\matrix{
+1\;\hfill&\hbox{if}\hfill&H_n>0,\hfill\cr
-1\hfill&\hbox{if}\hfill&H_n<0.\hfill
}\right.
\label{zerost}
\eeq
We start with two special cases where the analysis of ground states is
simpler.

\subsection{Directed model ($g=0$)}

The ground states of the directed model have been investigated
in~\cite{III,IV}.
In that case, the local field $H_n=h_n$ acting on grain $n$
only depends on the grains above $n$.
The expression~(\ref{zerost}) therefore boils down
to the following recursion relation:
\beq
\left\{\matrix{
h_n>0\;\lra\;\s_n=+1,\;\hfill&h_{n+1}=h_n-1,\hfill\cr
h_n<0\;\lra\;\s_n=-1,\;\hfill&h_{n+1}=h_n+\eps,\hfill
}\right.
\label{step}
\eeq
with initial values $h_1=0$, and $\s_1=+1$ for definiteness.
In a ground state, all the local fields $h_n$ lie in the range
\beq
-1\le h_n\le\eps.
\label{range}
\eeq
The boundedness of the local fields implies that
all the ground states have the same mean orientation $\mean{\s}$,
such that $\mean{f(\s)}=0$, hence
\beq
\mean{\s}=\frac{\eps-1}{\eps+1},
\label{aves}
\eeq
up to fluctuations which become negligible for large systems.

The number and the nature of ground states depend
on whether $\eps$ is rational or irrational.

If the shape parameter $\eps$ is irrational,
the recursion formula~(\ref{step}) implies that
all the local fields $h_n$ are non-zero (except $h_1=0$).
{\it A unique quasiperiodic ground state} is thus generated.
Had we made the initial choice $\s_1=-1$,
we would have obtained the same configuration,
up to a permutation of the two uppermost grains,
so that the model has in all two ground states.
For the golden mean~(\ref{gold}),
the ground-state grain configuration is given by a Fibonacci sequence:
\beq
\{\s_n\}=+--+--+-+--+--+-+--+-\cdots
\eeq

If the shape parameter $\eps$ is rational, i.e.,
\beq
\eps=\frac{p}{q}
\eeq
in irreducible form ($p$ and $q$ mutual primes),
some of the local fields $h_n$ generated by the recursion~(\ref{step}) vanish.
The corresponding grain orientations $\s_n$ remain unspecified.
This orientational indeterminacy occurs at points of perfect packing,
such that $n-1$ is a multiple of the {\it period} $p+q$.
The model therefore has {\it extensively degenerate ground states.}
Every one of them is a random sequence of two well-defined patterns of length
$p+q$, such that each pattern contains $p$ up and $q$ down grains.
Defining the static (configurational) entropy per grain as
\beq
\S=\frac{\ln\N_N}{N},
\label{ssdef}
\eeq
where $\N_N$ is the number of ground states of a system consisting of~$N$
grains, we have therefore
\beq
\S=\frac{\ln 2}{p+q}
\label{sdir}
\eeq
in the limit of a large system.
The simplest of all rational values is $\eps=1$, i.e., $p=q=1$.
In this symmetric case, we have $f(\s)=-\s$,
so that both orientations play symmetric r\^oles.
The ground states are all the {\it dimerised} configurations,
made of the patterns $+-$ and $-+$.
The static entropy here assumes its maximal value $\S=(\ln 2)/2$.

\subsection{Mean-field point ($g=1$, $\xii=\infty$)}

This situation is the complete opposite of the previous one.
Upper and lower grains have equal weights,
so that both components $h_n$ and $j_n$ of the local field add up to~give
\beq
H_n=\eta-f(\s_n),
\label{hmf}
\eeq
where we have introduced the {\it mean field}
\beq
\eta=\sum_{m=1}^Nf(\s_m)=\eps N^--N^+=\eps N-(\eps+1)N^+,
\label{etadef}
\eeq
with $N^+$ and $N^-=N-N^+$ being respectively the numbers of up grains and of
down grains in the configuration.
The above expression shows that $\eta$
is a {\it global} measure of excess void space~\cite{br} in the system.
This globality results from the exact cancellation
of fluctuations in the local void space corresponding to the competing
fields $h_n$ and $j_n$;
mean-field behaviour thus replaces the local fluctuations of the general case.
In this limit, the model resembles the one studied in~\cite{II},
one of the earliest building blocks of the present model.
We shall comment on further analogies between both models in due course.

The condition~(\ref{zerost}) thus reads
\beq
\s_n=\sign(\eta-f(\s_n)).
\eeq
It is fulfilled for all grains $n$
as soon as the mean field lies in the range $-1\le\eta\le\eps$.
These inequalities amount to saying that $N^+$ takes a well-defined
ground-state value:
\beq
N^+_\gs=\Int\left(\frac{\eps N+1}{\eps+1}\right),
\label{mgs}
\eeq
where $\Int(x)$, the integer part of $x$,
is the largest integer less than or equal to $x$.
The result~(\ref{aves}) is recovered in the limit of a large system.

At the mean-field point,
the ground states are all the configurations consisting
of $N^+_\gs$ up grains and $N^-_\gs=N-N^+_\gs$ down grains.
The number of ground states is therefore
\beq
\N_N=\bin{N}{N^+_\gs},
\label{nmf}
\eeq
so that the static entropy per grain reads
\beq
\S=\ln(\eps+1)-\frac{\eps}{\eps+1}\ln\eps.
\label{smf}
\eeq
This result will be illustrated in Figure~\ref{figss}.

\subsection{General case}

We now turn to the general case.
The behaviour of the model
turns out to be dictated mainly by the coupling constant $g$,
whereas the effect of the other parameters $\eps$ or $\xii$ is less
pronounced.
The difference between rational and irrational values of $\eps$,
which is the most salient feature of the directed model,
manifests itself most in the weak-coupling and strong-coupling regimes
($g\ll1$ and $g\gg1$).
The overall picture, already sketched in~\cite{V}, is the following.

If the shape parameter $\eps$ is irrational,
the static entropy rises continuously from zero
and behaves linearly at weak coupling:
\beq
\S\approx Ag\quad(g\ll1),
\eeq
where the amplitude $A$ depends on $\eps$ and $\xii$.
The meaning of this result~\cite{V}
is that generic ground states consist of quasiperiodic patches
whose typical length diverges as $\L(g)\sim1/g$ at weak coupling.
The entropy then smoothly increases as a function of $g$,
reaches a maximum around the mean-field coupling $g=1$,
and smoothly falls off to zero for $g\gg1$.

If the shape parameter $\eps$ is rational,
the static entropy stays equal to its value~(\ref{sdir})
over a whole range $0\le g\le g_\st$,
where $g_\st$ (with `s' for static) is the {\it static threshold}.
As already underlined in~\cite{V},
evaluating $g_\st$ is a non-trivial task in general.
The simplest situation is for $\eps=1$ and an even number of grains ($N=2K$).
The ground states in the directed case ($g=0$),
and by continuity at weak enough coupling, are the $2^K$ dimerised ones.
It can then be argued, thinking along the lines of~\cite{V},
that the first non-dimerised ground states which appear upon increasing $g$
are $++--(-+)^{K-2}$ and $--++(+-)^{K-2}$,
and that the relevant grain orientation for their stability
is the second one $(\s_2)$.
Consider the first configuration for concreteness.
We have $h_2=-1$
and $j_2=\exi+\exi^2+\exi^3-\exi^4+\cdots+\exi^{2K-3}-\exi^{2K-2}$.
The above configurations become ground states for $H_2=h_2+gj_2>0$, i.e.,
$g>g_\st=\abs{h_2}/j_2$.
We thus obtain
\beq
g_\st=\frac{1+\exi}{\exi(1+2\exi+2\exi^2-\exi^{N-2})},
\label{gsdef}
\eeq
and for an infinitely large system
\beq
g_\st=\frac{1+\exi}{\exi(1+2\exi+2\exi^2)}.
\label{gs}
\eeq

\begin{figure}[!ht]
\begin{center}
\includegraphics[angle=90,width=.65\linewidth]{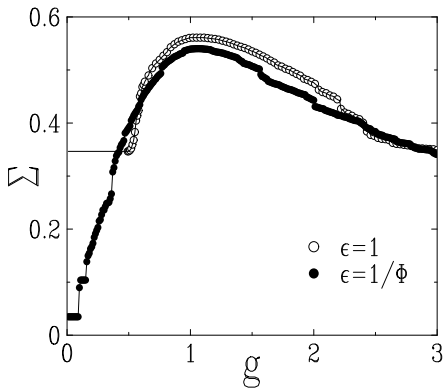}
\caption{\label{figent}
Plot of the static entropy $\S$ against the coupling constant $g$
for a column of $N=20$ grains with $\xii=10$
for both shape parameters $\eps=1$ and $\eps=1/\Phi$.
For $\eps=1$, the first symbol is at the threshold $g_\st\approx0.491651$
(see~(\ref{gsdef})).}
\end{center}
\end{figure}

Figure~\ref{figent} shows a plot of the static entropy $\S$
against~$g$ for a column of $N=20$ grains with $\xii=10$,
for both shape parameters $\eps=1$ and $\eps=1/\Phi$.
The exact number of ground states is obtained by means
of a full enumeration of the $2^{20}$ configurations.
The distinction between rational and irrational $\eps$
is visible at weak coupling.
For $\eps=1$ (our prototype of a rational),
$\S$ remains equal to $\S=(\ln 2)/2\approx0.346573$
in the whole range $0\le g\le g_\st$,
where the static threshold reads $g_\st\approx0.491651$,
including the finite-size effect of~(\ref{gsdef}).
For $\eps=1/\Phi$ (our prototype of an irrational),
$\S$ rises continuously from the minimal value $(\ln 2)/N$,
i.e., essentially zero, up to a finite-size effect
due to the existence of two ground states.

Besides this,
the static entropy has a weak dependence on the shape parameter $\eps$.
Both for rational and irrational~$\eps$,
the entropy has a smooth maximum around the mean-field point ($g=1$),
and it falls off smoothly at large~$g$.
The increase in the static entropy as the coupling~$g$ increases from 0 to 1
reflects the progressive decrease of the amount of order in the ground states.
For an irrational~$\eps$, this corresponds to a shortening
of the coherence length $\L(g)$ with increasing~$g$.
Finally, the weak dependence of the entropy on $\eps$ near its maximum,
i.e., near the mean-field point,
is an early indication that shape dependence is increasingly lost
as the model become more and more mean-field-like.

\subsection*{A digression on the effects of shape}

As mentioned above, increasing the value of $g$ corresponds to increasing the
frustration.
As soon as the coupling constant $g$ takes appreciable values,
the evolution of ordering due to compaction no longer proceeds in a
top-down fashion, as it did in the directed ($g=0$) model~\cite{II,III,IV}.
We outline here what we might expect for the dynamical behaviour in the case of
general $g$, before making specific calculations.

Looking at zero-temperature dynamics allows us to get a flavour of
the ordering behaviour.
Do grains retrieve their ground states in the limit of zero perturbation, and
if so, how?
In the case of $g=0$, the effect of the shape parameter~$\eps$ is maximal
on the zero-temperature dynamics~\cite{II,III,IV}:
irregularly shaped grains with irrational $\eps$ order ballistically fast
into their unique quasiperiodic ground state, while regularly shaped grains
never retrieve any of their many ground states,
manifesting density fluctuations instead,
due to the presence of many sites where $h_n=0$~\cite{III,IV}.
Similar behaviour has been observed in experiments~\cite{sid} on
vibrated granular beds,
from which we can infer that the effects of frustration, if any, are
negligible compared to the intensity of vibration
in the experiment under consideration.

For a low but non-zero coupling $g$,
the `reverse' field $j_n$ induces some frustration
as it begins to impose its own order in an upward direction,
competing with the downward ordering due to $h_n$.
In the case of irrational $\eps$,
one might imagine that the effect of the reverse ordering would break up the
unique quasiperiodic ground state corresponding to $g=0$;
while in the case of rational $\eps$,
even the tiniest amount of frustration ensures that points of zero field are
not constantly generated and re-generated,
and hence that the density fluctuations of the $g=0$ case disappear.
The only possible logical culmination
of ordering from {\it both} the top and the bottom of the column might
be to 'fix' the points of zero field; in the case of $\eps=1$ this corresponds
to a dimerisation of the ground states.
These were indeed the findings of earlier work~\cite{V}.

Increasing frustration beyond these values might in a general sense lead
to the breakdown of even this level of order.
While the details would depend on parameters like~$\xii$ and $\xidy$,
we mention a few likely outcomes.
For both weak ($g\ll1$) and extremely strong ($g\gg1$) couplings,
we might expect the prevalence of very similar order,
although propagating in opposite directions
(top-down in the first, and upwards in the second case).
At the mean-field point ($g=1$, $\xii=\xidy=\infty$),
where the fluctuations in both ordering fields cancel each other,
the system of grains is totally uncorrelated,
leading to a situation similar to that explored in~\cite{II}.
The gradual replacement of the ordering by individual grains by the
ordering of granular clusters might be expected to occur for $g$ in the
vicinity of the mean-field point ($g=1$);
shape effects are therefore expected to be minimal here.

The results of the next two sections will bear out some of these
speculations.

\section{Dynamics: jamming time}
\setcounter{equation}{0}
\def\theequation{4.\arabic{equation}}

The rules for zero-temperature dynamics
are defined as follows~\cite{III,IV,V}.
The uppermost grain is kept fixed to
\beq
\s_1=+1.
\label{upper}
\eeq
The other grains ($n=2,\dots,N$) are selected at a rate given by~(\ref{odef}).
Once a grain is selected,
its orientation~$\s_n$ is aligned along the local field $H_n$
according to the deterministic rule
\beq
\s_n\to\sign H_n,
\label{zerodyn}
\eeq
provided the local field $H_n$ does not vanish.
The choice of boundary condition~(\ref{upper}) is motivated
by the fact that the strongest component
in the local field $H_n$ is typically the long-ranged component $h_n$
which propagates via gravity.
As a consequence, interactions propagate downwards in general,
so that it is natural to impose a boundary condition at the top of the column.
We assume that the column is prepared in a random state,
where each grain is oriented at random
($\s_n=\pm1$ with equal probabilities for all $n\ge2$).

The above rule is well-defined for a non-zero coupling constant $g$,
because the local fields $H_n$ do not vanish in general.
The zero-temperature dynamics thus defined leads to {\it metastability}.
A finite column of $N$ grains is eventually driven
to an absorbing configuration or {\it attractor},
in a finite {\it jamming time} $T$.
This attractor is necessarily one of the ground states described earlier,
i.e., a configuration where every orientation $\s_n$ is aligned with $H_n$.
Let us emphasize that imposing a restrictive boundary condition,
i.e., fixing one of the spins (see~(\ref{upper})),
is necessary to have metastability in the above sense.
Without such restriction,
zero-temperature dynamics would drive a finite system
to a fluctuating steady state.

In the present context, `metastable state', `attractor' and `ground state'
are therefore essentially synonymous.
Arbitrary initial conditions can lead to any one
of the ground states being reached.
They are however {\it fragile}, in the sense that a slightly different
initial condition or stochastic history leads to another
attractor being reached in general.
This fragility~\cite{cates} of metastable states is one of the characteristics
of granular media~\cite{0}.

Along the lines of~\cite{V},
we will focus on two aspects of zero-temperature dynamics,
namely the statistics of the jamming time (in this Section)
and that of the attractors (in Section~5).
The jamming time is doubly random,
as it depends both on the initial configuration of the system
and on its whole stochastic history.
Just as for statics, we begin with a few special cases.

\subsection{Directed model ($g=0$)}

The dynamical behaviour of the directed column
has been studied at length~\cite{III,IV}.
It depends qualitatively on whether $\eps$ is rational or irrational.

If the shape parameter $\eps$ is irrational,
its unique quasiperiodic ground state is reached by ballistic coarsening.
An upper layer of the column is ordered,
whose thickness grows linearly with time
(the ballistic phase will be described more thoroughly in Section 4.3).
The dependence of the corresponding velocity $V$ on $\eps$
has been investigated in~\cite{IV}.

If the shape parameter $\eps=p/q$ is rational, the local field $h_n$ may vanish
whenever $n-1$ is a multiple of the period $p+q$.
It is therefore natural to complete the dynamical rule~(\ref{zerodyn})
as~\cite{III,IV}:
\beq
\s_n\to\left\{\matrix{
+\hfill&\hbox{if}\hfill&h_n>0,\cr
\pm\;\hfill\hbox{ with prob.~}1/2\hfill&\hbox{if}\hfill&h_n=0,\cr
-\hfill&\hbox{if}\hfill&h_n<0.\cr
}\right.
\label{zerodynrat}
\eeq
These dynamical rules do not drive the system to any of its ground states.
There are always grains whose local fields~$h_n$ vanish.
The column reaches a non-trivial fluctuating steady state,
investigated in~\cite{IV},
which exhibits anomalous roughening: the fluctuations in the local field
grow with a power law, as $\mean{h_n^2}\sim n^{2/3}$.

\subsection{Mean-field point ($g=1$, $\xii=\infty$)}

It has been shown in Section 3.2 that the statics of the model
is of a mean-field type when $g=1$ and $\xii=\infty$.

This property extends to the dynamics in the $\xidy=\infty$ limit,
where activation energies are negligible, so that grains are sampled uniformly
and the effect of gravity is lost.
In this limit, the dynamical rule~(\ref{zerodyn}) indeed becomes
\beq
\s_n\to\sign(\eta-f(\s_n)),
\label{mfdyn}
\eeq
where $\eta$ is the mean field introduced in~(\ref{etadef}).
Thus every grain orientation $\s_n$ is updated to $+1$ (resp.~$-1$),
with unit rate, irrespective of its position $n$,
as long as $\eta>\eps$ (resp.~$\eta<-1$).
This rule can be recast as the following effective dynamics
for the number $N^+$ of up spins:
\beq
\left\{\matrix{
N^+<N^+_\gs\;\lra\;N^+\to N^++1\hfill&\mbox{with rate}&N-N^+,\hfill\cr
N^+>N^+_\gs\;\lra\;N^+\to N^+-1\hfill&\mbox{with rate}&N^+.\hfill
}\right.
\eeq
The dynamics stop as soon as $N^+$ reaches the value $N^+_\gs$
(see~(\ref{mgs})), i.e., when the system reaches a ground state.

Mean-field zero-temperature dynamics are fast,
in the strong sense that the jamming time is microscopic.
More precisely, consider $\eps<1$ for definiteness, so that $N^+_\gs<N/2$.
For a random initial configuration, characterised by $N^+\approx N/2$,
the mean jamming time can be shown to read (see e.g.~\cite{redner})
\beq
\mean{T}\approx\sum_{N^+=N^+_\gs}^{N/2}\frac{1}{N^+}
\approx\ln\frac{\eps+1}{2\eps}
\label{mfpt}
\eeq
for a large system.
The jamming time is indeed found to be microscopic.
This makes good physical sense, since the system is fully uncorrelated,
and all the grains are simultaneously mobile.

Furthermore, as $N^+$ is the only non-trivial dynamical variable,
it is clear that all the ground states are reached with uniform probability.
In other words, anticipating the discussion of Section~5,
the mean-field dynamics of our column model
($g=1$, $\xii=\xidy=\infty$)
is one of the rare instances where Edwards' flatness hypothesis~\cite{edwards}
can be shown to be exactly valid.
A similar result has been established
in the context of the ageing dynamics of mean-field spin-glass models~\cite{fv}.

\subsection*{R\^ole of a finite $\xidy$}

We now study the model at its static mean-field point,
but with generic dynamics defined by a finite value
of the dynamical length $\xidy$.
This situation is of interest because it is both simple
(the updating rule is still given by~(\ref{mfdyn})),
and non-trivial (grains are not selected uniformly anymore).
Grain $n$ is indeed updated at a rate given by~(\ref{odef}),
so that upper grains are more mobile than lower ones.
This case is physically similar to that of a column of
non-interacting grains in the presence of gravity~\cite{II}.

Consider again $\eps<1$ for concreteness.
For a random initial configuration,
the typical number of grains to be flipped from $+$ to $-$ reads
\beq
N_\f\approx\frac{N}{2}-N^+_\gs\approx\frac{1-\eps}{2(1+\eps)}\,N.
\eeq

It is worth considering first the slow regime ($N\gg\xidy$).
In this situation, the grains to be flipped are essentially
the $N_\f$ uppermost $+$ grains,
which occupy an upper layer of depth $2N_\f$ in a random initial state.
As a consequence, the jamming time scales as $T\sim\exp(2N_\f/\xidy)$, i.e.,
\beq
T\sim\exp\left(\frac{1-\eps}{1+\eps}\frac{N}{\xidy}\right).
\label{tmf}
\eeq
The jamming time therefore grows exponentially with $N$.
The prefactor is proportional to that entering
the asymptotic orientation~(\ref{aves}).
As a consequence,~(\ref{tmf}) does not apply to the symmetric situation ($\eps=1$),
where the jamming time is not exponentially large in $N$ in this regime.
Large jamming times testify that the retrieval of ground states is not easy;
they are likely to correspond to lower dynamical entropies.
The $\eps$-dependent
prefactor suggests that shape effects would be strongly related
to the absence of Edwards flatness, as will be verified below.
Furthermore, the typical orientation profile of an attractor
takes the form of a discontinuous step:
\beq
\mean{\s_n}\approx\left\{\matrix{
-1\hfill&(n<2N_\f),\hfill\cr
0\hfill&(2N_\f<n<N).\hfill}\right.
\label{slowpro}
\eeq

\begin{figure}[!ht]
\begin{center}
\includegraphics[angle=90,width=.65\linewidth]{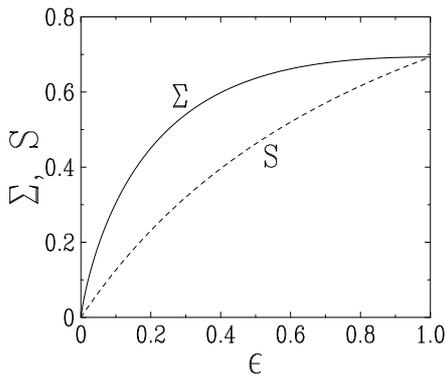}
\caption{\label{figss}
Comparison between static and dynamical zero-temperature entropies
against the shape parameter $\eps$ in the range $0\le\eps\le1$,
in the slow regime of the mean-field model ($g=1$, $\xii=\infty$,
$N\gg\xidy$).
Upper full curve: static entropy~$\S$ (see~(\ref{smf})).
Lower dashed curve: dynamical entropy~$S$ (see~(\ref{sd})).}
\end{center}
\end{figure}

We define the zero-temperature dynamical entropy per grain as
\beq
S=-\frac{1}{N}\sum_\C Q(\C)\ln Q(\C),
\label{sddef}
\eeq
where the sum runs over the attractors $\C$,
and where $Q(\C)$ is the probability that the dynamics drive the system
into attractor number $\C$, starting from a random initial condition.
The above picture of jamming in the slow regime leads to
the estimate $NS\approx(N-2N_\f)\ln 2$,
as the difference $N-2N_\f$ is an estimate of the number
of the lower grains which do not move during the history of the column.
The dynamical entropy therefore reads
\beq
S=\frac{2\eps}{1+\eps}\,\ln 2.
\label{sd}
\eeq
This quantity is smaller than the static entropy $\S$, given by~(\ref{smf}).
Both entropies indeed only coincide at the extremal values $\eps=0$
(where $\S=S=0$) and $\eps=1$ (where $\S=S=\ln 2$).
The entropy difference is maximal for $\eps=1/4$,
where it equals $\S-S=\ln(5/4)\approx0.223143$.
Figure~\ref{figss} presents a comparison between both entropies.
This slow mean-field regime is one of the rare cases
where the violation of Edwards' flatness
can be turned into a quantitative estimate.
Another zero-temperature example is provided
by kinetically constrained one-dimensional spin models~\cite{gds}.

\begin{figure}[!ht]
\begin{center}
\includegraphics[angle=90,width=.65\linewidth]{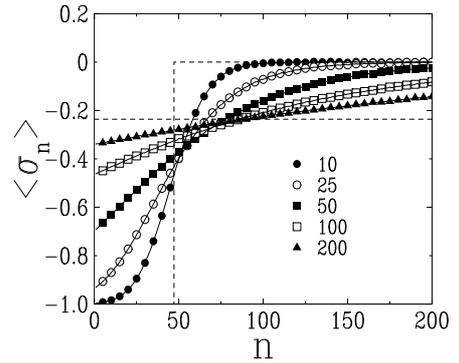}
\caption{\label{figpro}
Plot of the orientation profile $\mean{\s_n}$
of the attractors against depth $n$
at the static mean-field point ($g=1$, $\xii=\infty$)
for $\eps=1/\Phi$ and $N=200$.
Symbols: data for several values of~$\xidy$.
Dashed lines: limiting uniform and step profiles,
respectively corresponding to $N\ll\xidy$ and $N\gg\xidy$.}
\end{center}
\end{figure}

For generic values of the ratio $N/\xidy$,
the attractor statistics vary continuously between the uniform case
of mean-field dynamics (for $N\ll\xidy$)
and the non-uniform case of the slow regime,
described above (for $N\gg\xidy$).
This continuous dependence is best visualised by the orientation profile
$\mean{\s_n}$ of the attractors.
Figure~\ref{figpro} shows numerical data for the orientation profile
with $\eps=1/\Phi$ and $N=200$.
Each dataset is obtained by averaging
over $10^6$ different stochastic histories
with different initial configurations.
The data demonstrate a continuous crossover
between a uniform profile
at the mean value~(\ref{aves}), i.e., $\mean{\s}=-1/\Phi^3\approx-0.236067$
(for $N\ll\xidy$)
and the discontinuous step profile~(\ref{slowpro})
of the slow regime (for $N\gg\xidy$).

\subsection{Full dynamical phase diagram}

We now turn to the zero-temperature dynamics of our model
for generic parameter values.
We expect that the phase diagram of the model can be roughly divided into three regions:

\noindent (a) the weak-coupling regime ($g\ll1$),
already explored in earlier work for $\eps=1$~\cite{V}; a strong
uniform field $h_n$ is the dominant effect, leading to a correspondingly
strong dependence on the shape parameter $\eps$,

\noindent (b) the neighbourhood of the mean-field point ($g=1$),
where ordering proceeds with fewer local constraints, so that shape dependence
is increasingly lost,

\noindent (c) the strong-coupling end ($g\gg1$), where a strong
frustrating field $j_n$ is the dominant effect; this might be expected to
lead to the return of a strong dependence on the shape parameter $\eps$.

\begin{figure}[!ht]
\begin{center}
\includegraphics[angle=90,width=.475\linewidth]{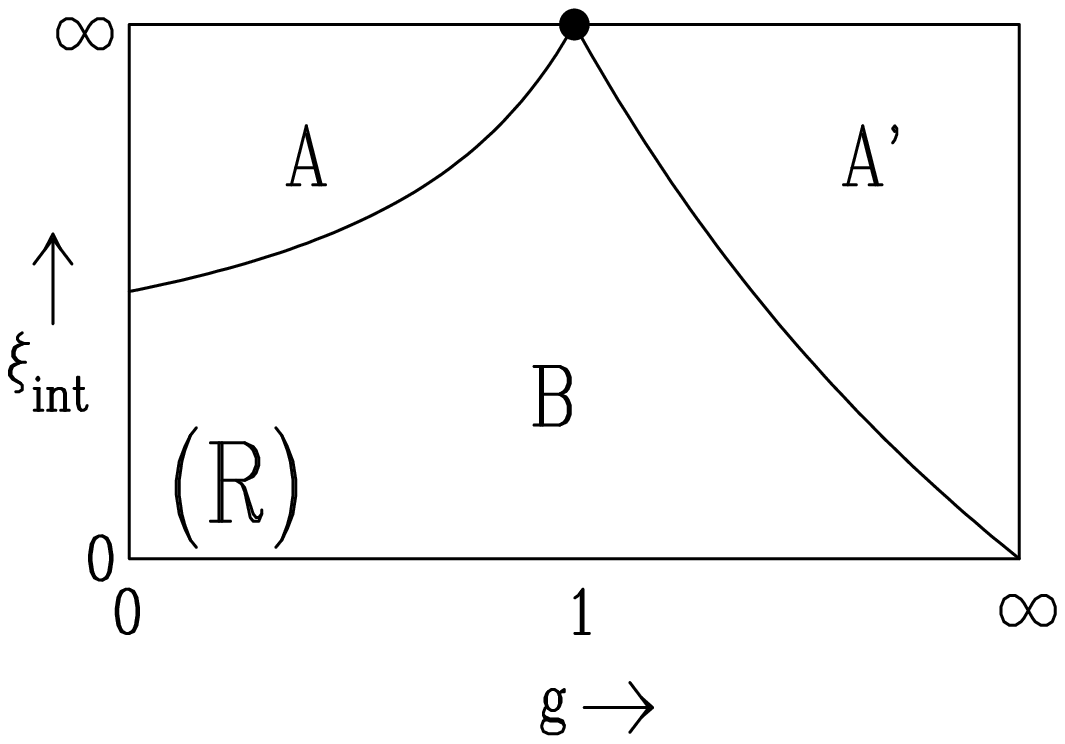}
\includegraphics[angle=90,width=.475\linewidth]{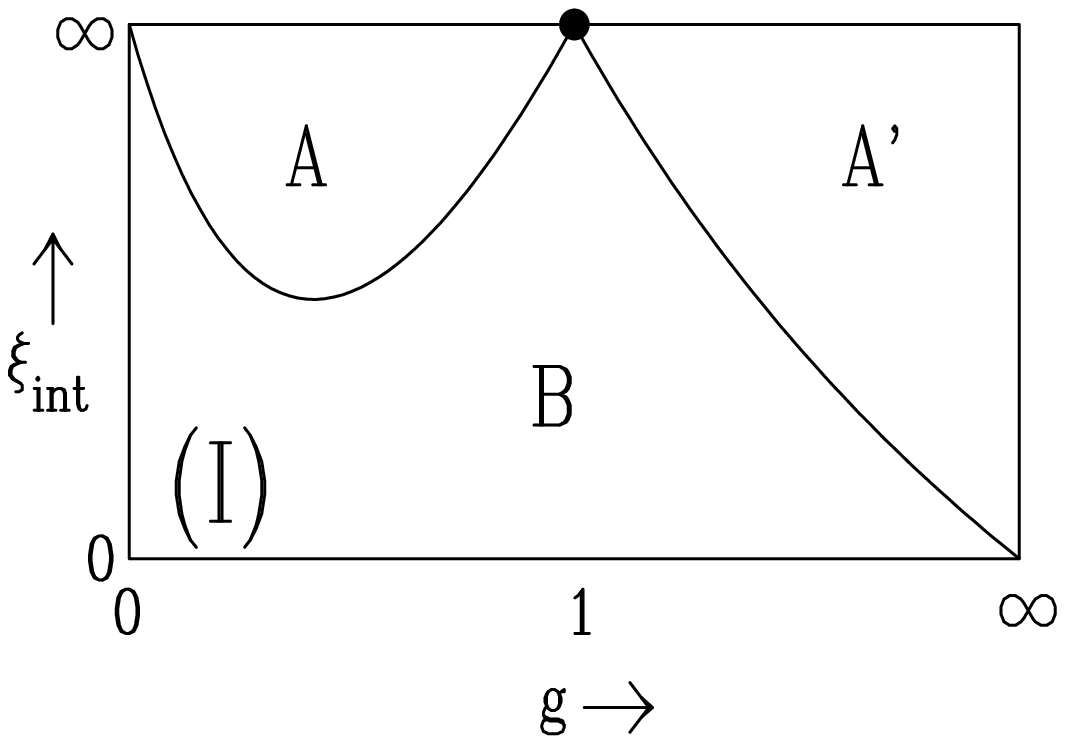}
\caption{\label{figdiag}
Sketch of the phase diagram of the model in the $g$--$\xii$ plane
for a rational ((R), left) and irrational ((I), right) shape parameter $\eps$.
Symbol: mean-field point.
Curves: critical lines $\xii=\xiic(g;\eps)$.
B: ballistic phase.
A: weak-coupling activated phase.
A': strong-coupling activated phase.}
\end{center}
\end{figure}

For the time being, we restrict our study to the case where $\xidy=\infty$,
so that grains are sampled uniformly by the dynamics.
The dynamical properties of the model are mainly dictated
by the coupling constant $g$ and the interaction length $\xii$,
with a less pronounced dependence on the shape parameter $\eps$,
except in the weak-coupling regime.
The main features of the model are summarised
in the dynamical phase diagram shown in Figure~\ref{figdiag}.
This picture will be made more precise in the following.

At this point, it is worth emphasising
that the dynamical phase diagram of the model depends on the precise
definition of zero-temperature dynamics,
including the boundary condition~(\ref{upper}).
In particular,
had we chosen to fix the bottommost spin ($\s_N$) instead of the uppermost one
($\s_1$),
we would have obtained somewhat different phase boundaries; this would be
true especially for large $g$,
with a finite limiting $\xiic$ along the $g=\infty$ axis mirroring
that which exists currently for $g=0$.
We will return to this point in Section~6.

The scenario observed for generic values of $g$
qualitatively follows that of the weak-coupling ($g\ll1$) regime for $\eps=1$,
investigated in~\cite{V}.
The model is in a ballistic phase if the interaction length is small
($\xii<\xiic$),
and in an activated phase if the interaction length is large ($\xii>\xiic$).
The critical value $\xiic$ of the interaction length $\xii$
depends strongly on $g$ and weakly on $\eps$.
Let us start by reviewing the main characteristics of both phases
and of the crossover between them, which have been analyzed in~\cite{V}.

\noindent$\bullet$ {\it Ballistic phase ($\xii<\xiic$).}
In this phase, zero-tempe\-ra\-ture dynamics propagate order into the
system from the top down~\cite{III,IV,V}.
More precisely, if we define the thickness~$L(t)$
of the upper ordered layer of the column as the depth
of the uppermost grain which is {\it not} aligned with its local field,
the ballistic phase is characterised by a linear growth of the mean thickness:
\beq
\mean{L(t)}\approx Vt,
\label{vt}
\eeq
where $V$ is the ballistic velocity.
Fluctuations around this mean behaviour are due to diffusion.
As a consequence, for a finite system of $N$ grains,
the jamming time grows linearly with $N$:
\beq
\mean{T}\approx\frac{N}{V},
\label{tlin}
\eeq
up to relatively negligible fluctuations,
so that the reduced variance of the jamming time,
\beq
K_T=\frac{\var{T}}{\mean{T}^2}=\frac{\mean{T^2}}{\mean{T}^2}-1,
\label{kdef}
\eeq
is of order $1/N$.

\noindent$\bullet$ {\it Activated phase ($\xii>\xiic$).}
In this phase,
zero-tempe\-ra\-ture dynamics do not proceed in any ordered way.
The system explores its configuration space more or less uniformly,
until it meets one of its ground states by chance.
This picture is that of an activated phenomenon.
An exponential growth of the mean jamming time with the column size
results:
\beq
\mean{T}\sim\e^{aN},
\label{tact}
\eeq
at least for very large $N$,
where $a$ is the effective reduced activation energy per grain.
In other words, the system has to cross an extensive entropic barrier,
whose height grows asymptotically as $aN$, in order to reach a ground state.
This also suggests an exponential distribution of jamming times,
so that the reduced variance asymptotes to $K_T=1$.

\noindent$\bullet$ {\it Diffusive crossover ($\xii\approx\xiic$).}
The crossover between ballistic and activated behaviour
has been shown~\cite{V} to be described by a simple effective model.
The thickness~$L(t)$ of the ordered layer was treated as a collective coordinate,
and its dynamics modelled by biased Brownian motion
(see Appendix B of~\cite{V}).
In the ballistic regime, the downward propagation of the
layer is helped by the dominant effect of the field $h_n$, while in the
activated regime, it is hindered by the dominant effect of the field $j_n$.
The crossover thus corresponds to the point
$\xii=\xiic$ where the effects of the two fields are neutralised.
The behaviour right at the crossover is dictated by
the presence of a diffusive critical point.
Observables obey finite-size scaling laws
involving the scaling variable $z=\alpha X+\beta$, with $X=N(\xii-\xiic)$,
whereas $\alpha $ and $\beta$ are non-universal numbers.
For instance, the mean jamming time $\mean{T}$
and its reduced variance $K_T$ obey
\beq
\mean{T}\approx\frac{N^2}{2D}\,F(z),\quad K_T\approx G(z),
\label{fss}
\eeq
where $D$ is an effective diffusion constant,
whereas $F$ and $G$ are known universal scaling functions.
The reduced variance of the jamming time takes the non-trivial universal value
$K_T=G(0)=2/3$ right at the critical point,
i.e., for $z=0$ or $\xii\approx\xiic-\beta/(\alpha N)$.
In practice the $1/N$ correction is negligible for $N/\xiic>10$.
Measuring $K_T$ thus provides an efficient way
of exploring the dynamical phase diagram.

\begin{figure}[!ht]
\begin{center}
\includegraphics[angle=90,width=.65\linewidth]{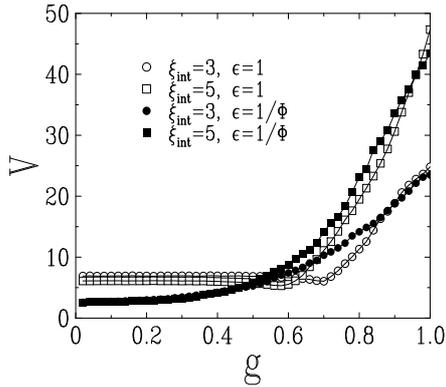}
\caption{\label{figv}
Plot of the ballistic velocity $V$ against the coupling constant $g$
for several values of the parameters $\xii$ and $\eps$.}
\end{center}
\end{figure}

We now turn to the presentation and discussion of actual numerical data.
Figure~\ref{figv} shows a plot of the ballistic velocity $V$
against the coupling constant $g$,
for both shape parameters $\eps=1$ and $\eps=1/\Phi$,
and for two values of the interaction length ($\xii=3$ and 5)
deep in the ballistic phase, i.e., much smaller than $\xiic$.
For the irrational shape parameter $\eps=1/\Phi$,
the velocity $V$ departs continuously from its value
in the directed model ($g=0$), i.e., $V\approx2.58$~\cite{IV}.
It increases steadily and reaches its maximal value
at or near the mean-field coupling ($g=1$).
If the shape parameter $\eps$ is rational,
the velocity assumes a constant value
over a whole range $0\le g\le g_\dy$,
where $g_\dy$ (with `d' for dynamical) is the {\it dynamical threshold}.
Along the lines of~\cite{V}, the latter can be shown to read
\beq
g_\dy=\frac{1-\exi}{\exi\;{\rm max}(p,q)}=\frac{\e^{1/\xii}-1}{{\rm max}(p,q)}
\label{gdgene}
\eeq
for an arbitrary rational value $\eps=p/q$ of the shape parameter.
For $\eps=1$, this result reads
\beq
g_\dy=\frac{1-\exi}{\exi}=\e^{1/\xii}-1.
\label{gd}
\eeq
This dynamical threshold is always smaller
than its static counterpart (see~(\ref{gs})), as it should be.
Both thresholds will be shown in Figure~\ref{figlines}.

The {\it increase} of $V$ with $g$ to a maximum near the mean-field point,
observed for both irrational and rational~$\eps$, seems apparently
paradoxical, as increasing $g$ after all increases the effects of frustration.
The dynamics at the mean-field static point ($g=1$) are
however expected to be rather fast, irrespective of shape,
so that the velocity can be expected to be both large and independent of
$\eps$ near the mean-field point.
We will return to this paradox later,
when we investigate the scaling properties of the jamming time
near the mean-field point.

\begin{figure}[!ht]
\begin{center}
\includegraphics[angle=90,width=.65\linewidth]{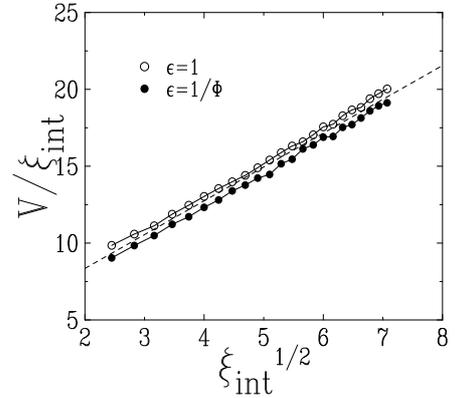}
\caption{\label{figvc}
Plot of the ballistic velocity $V$ at the mean-field coupling ($g=1$),
divided by the interaction length $\xii$, against $\xii^{1/2}$,
for both shape parameters $\eps=1$ and $\eps=1/\Phi$.
The dashed line has slope 2.2.}
\end{center}
\end{figure}

Right at the mean-field coupling ($g=1$), the ballistic velocity $V$
is observed to depend strongly on $\xii$ and weakly on $\eps$,
along the lines of the above discussion.
It diverges faster than linearly at large $\xii$.
In order to investigate this divergence, we present in Figure~\ref{figvc}
a plot of the ratio $V/\xii$ against $\xii^{1/2}$,
for both shape parameters $\eps=1$ and $\eps=1/\Phi$.
The largest velocities reached are $V\approx1000$ for $\xii=50$; velocities
larger than this are hard to measure accurately.
The observed linear behaviour of both datasets with equal slopes
suggest the scaling behaviour
\beq
V(g=1)\approx A\,\xii^{3/2},
\label{vc}
\eeq
the prefactor $A\approx2.2$ being seemingly independent of shape
parameter.
The exponent 3/2 of the growth of the velocity,
as mean-field dynamics are approached,
can be justified heuristically as follows.
First of all, the dynamics are expected to proceed
by ordering not individual grains,
but entire correlated clusters of typical length $\xii$.
This already brings in a factor of $\xii$.
Furthermore, if we consider one such piece
after a microscopic time but before ordering,
we see that typical fluctuations of the mean orientation around
its ground-state value~(\ref{mgs}) are expected to fall off as $\xii^{-1/2}$
(because of the law of large numbers).
The formula~(\ref{mfpt}) shows that the corresponding times
also fall off as $\xii^{-1/2}$.
This brings in an extra factor of $\xii^{1/2}$ to the velocity,
providing a plausible mechanism for the exponent 3/2 in~(\ref{vc}).

\begin{figure}[!ht]
\begin{center}
\includegraphics[angle=90,width=.65\linewidth]{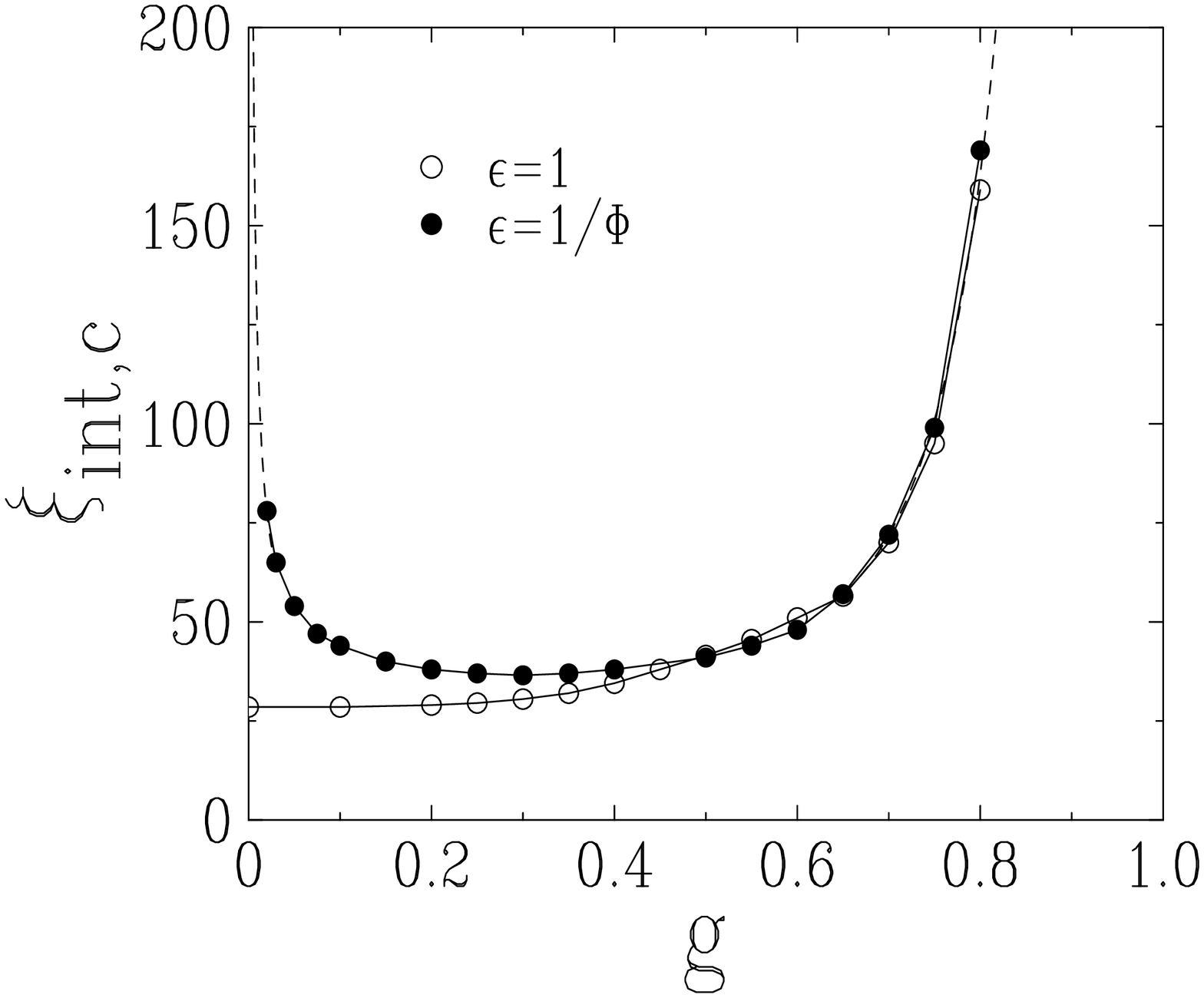}

\includegraphics[angle=90,width=.65\linewidth]{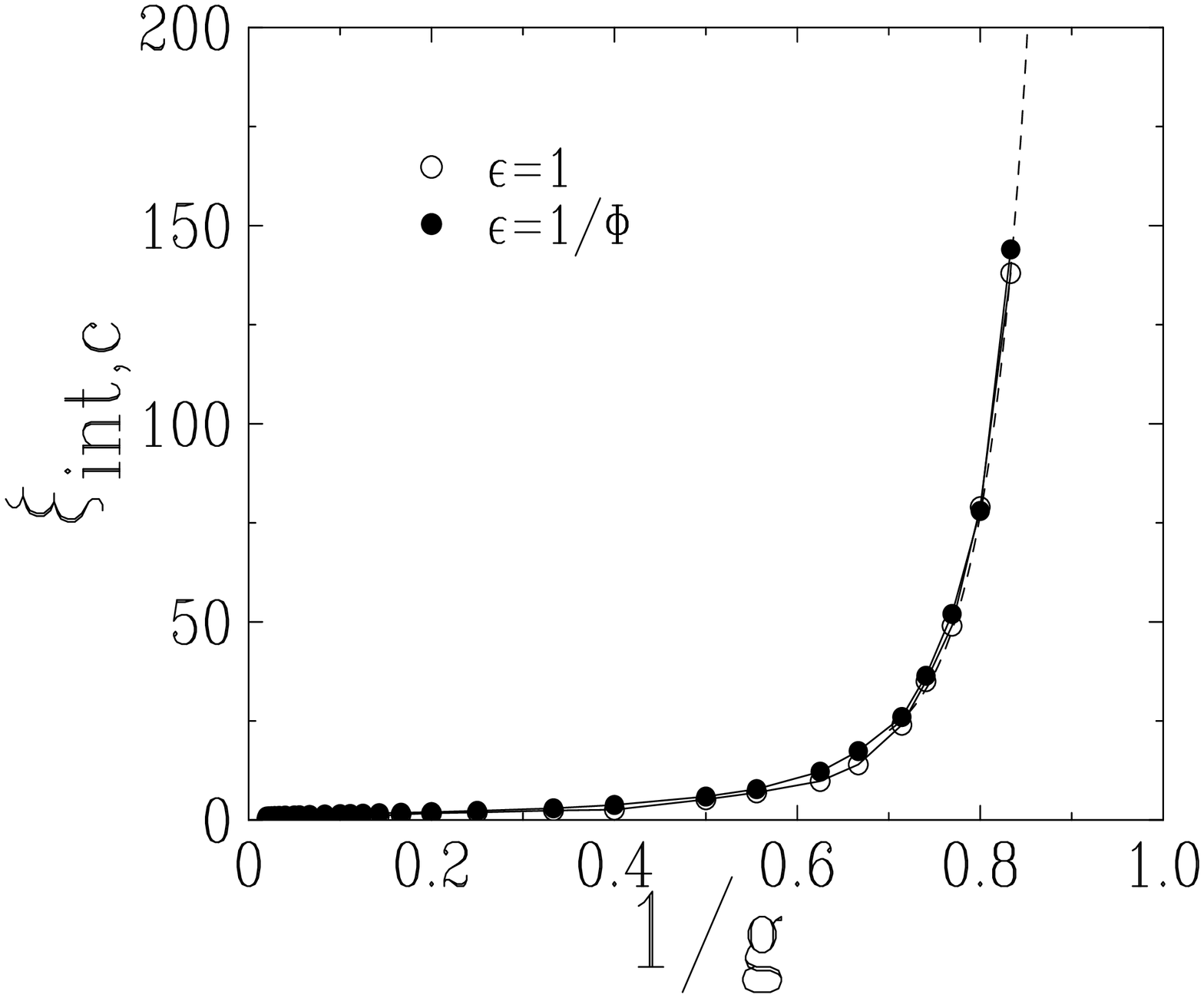}
\caption{\label{figxc}
Plot of the critical lines
for both shape parameters $\eps=1$ and $\eps=1/\Phi$.
Top: $\xiic$ separating phases A and B against $g$ for $g<1$.
Bottom: $\xiic$ separating phases A' and B against $1/g$ for $g>1$.
Dashed lines: fits
incorporating the divergence laws~(\ref{xone}) with $B_1=11$
and~(\ref{xzero}) with $B_0=0.83$.}
\end{center}
\end{figure}

\begin{figure}[!ht]
\begin{center}
\includegraphics[angle=90,width=.65\linewidth]{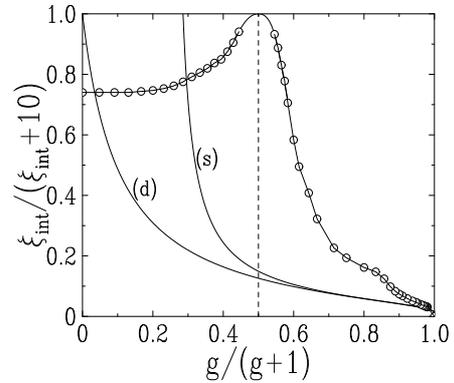}
\caption{\label{figlines}
Plot of notable lines in the $g$--$\xii$ plane for $\eps=1$,
providing a quantitative version of the left panel of Figure~\ref{figdiag}.
Coordinates have been chosen for the sake of clarity.
Left full line (d): dynamical threshold~(\ref{gd}).
Right full line (s): static threshold~(\ref{gs}).
Dashed line: mean-field coupling ($g=1$).
Line with symbols: critical line.}
\end{center}
\end{figure}

We now present quantitative data for the phase diagram
sketched in Figure~\ref{figdiag}.
The position of the critical line $\xiic$ has been
obtained by means of the criterion $K_T=2/3$
for large enough systems, as explained in the paragraph below~(\ref{fss}).
Figure~\ref{figxc} shows plots of $\xiic$ against the coupling constant
for both shape parameters $\eps=1$ and $\eps=1/\Phi$.
In practice, values of $\xiic$ larger than 100 or 200
become very hard to measure with sufficient accuracy.

The critical value $\xiic$ is observed to diverge as $g\to1$ from both sides,
consistent with the result that mean-field dynamics are fast.
The fits shown on Figure~\ref{figxc} as dashed lines
suggest a quadratic divergence of the form
\beq
\xiic\approx\frac{B_1}{(g-1)^2}
\label{xone}
\eeq
as mean-field coupling is approached from both sides.
The amplitude is estimated to be $B_1\approx11\pm3$, irrespective of $\eps$,
albeit with a relatively large uncertainty.
The above formula will be corroborated
below by the finite-size scaling law~(\ref{dipfss}).

The behaviour of $\xiic$ at weak coupling depends on whether $\eps$ is
rational or not.
For the rational shape parameter $\eps=1$,
the non-trivial value $\xiic\approx28.4$ at $g=0$~\cite{V} is recovered.
For the irrational shape parameter $\eps=1/\Phi$,~$\xiic$ is observed
to diverge as $g\to0$.
The fit shown as a dashed line suggests a linear divergence of the form
\beq
\xiic\approx\frac{B_0}{g},
\label{xzero}
\eeq
with amplitude $B_0\approx0.8$.
The above divergence can be explained in terms of the statics of the model.
The quasiperiodic ground state obtained for irregular grains at $g=0$
splits into patches of typical length $\L(g)\sim 1/g$ at small but non-zero
coupling.
As each of these quasiperiodic patches is retrieved coherently,
it is natural to expect that the critical interaction length $\xiic$
should diverge in proportion to the static length $\L(g)$ at weak coupling.

Since the critical interaction length $\xiic(g)$ diverges both as $g\to0$ and
$g\to1$ for an irrational shape parameter,
it must exhibit a minimum somewhere in the range $0<g<1$.
For $\eps=1/\Phi$, this minimal value $\xiic\approx36$,
reached for $g\approx0.3$,
is in the same ball park as the weak-coupling value $\xiic\approx28.4$ for
$\eps=1$.
Soon after this minimum is attained,
the values of $\xiic$ for $\eps=1$ and $\eps=1/\Phi$ begin to be nearly
identical.
One can therefore view the minimum in $\xiic$ as a crossover between a phase
where the dynamics of individual grains (strong shape dependence) governs the
retrieval of
weak-coupling-like ground states, and another where the dynamics of clusters
(little shape dependence) governs the retrieval of mean-field-like ground
states.
This viewpoint makes
it clear why the most pronounced effects of shape occur before the crossover.

In the case of the rational shape parameter $\eps=1$,
Figure~\ref{figlines} shows a comparison between
the critical line in the $g$--$\xii$ plane and
the static and dynamical thresholds~$g_\st$ and $g_\dy$,
respectively given by~(\ref{gs}) and~(\ref{gd}).
We recall that the dynamics is fully independent of $g$
below the dynamical threshold ($0<g<g_\dy$),
whereas the attractors are exactly the dimerised configurations
below the (larger) static threshold ($0<g<g_\st$).
It turns out that neither threshold has any effect on the fast (ballistic)
or slow (activated) nature of the dynamics.

\begin{figure}[!ht]
\begin{center}
\includegraphics[angle=90,width=.65\linewidth]{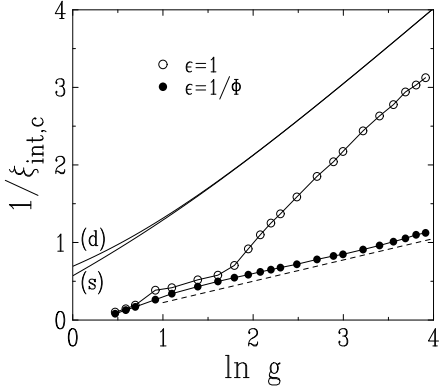}
\caption{\label{figxdet}
Plot of $1/\xiic$ against $\ln g$,
for both shape parameters $\eps=1$ and $\eps=1/\Phi$.
The dashed line has slope 0.275.
Upper full lines: static (s) and dynamical (d) thresholds for $\eps=1$.}
\end{center}
\end{figure}

The three lines shown in Figure~\ref{figlines} seem to become close to each
other in the lower right corner of the plot, i.e., at strong coupling,
suggesting that one should look more closely at this regime.
For large values of $g$, the dominant effect is that of the $j_n$ field
which propagates interactions upwards along the column.
As a consequence, with our choice of fixing the uppermost spin
(see~(\ref{upper})),
the system will find it more and more difficult to order.
This explains in qualitative terms why the critical line $\xiic$ falls off
to zero at large $g$.
The data shown in Figure~\ref{figxdet} suggest an inverse logarithmic law for
$\xiic$, of the form
\beq
\xiic\approx\frac{B_\infty}{\ln g}.
\label{xl}
\eeq
For the irrational shape parameter $\eps=1/\Phi$,
the data show a smooth linear growth with slope $1/B_\infty\approx0.275$,
i.e., $B_\infty\approx3.6$.
For the rational shape parameter $\eps$,
the data are observed at first to follow those for $\eps=1/\Phi$,
and then to cross over rather abruptly to a steeper regime of growth.
The data seem to become asymptotically
parallel to the static and dynamical thresholds $g_\st$ and $g_\dy$.
If this observation holds quantitatively, we obtain
the asymptotic slope $B_\infty=1$ for $\eps=1$.

\subsection*{Vicinity of the mean-field point}

We now turn to the study of the apparently paradoxical dependence of
the jamming time $T$ on the interaction length,
in the vicinity of the mean-field point.
We have encountered another avatar of the same paradox earlier,
concerning the behaviour~(\ref{vc}) of the velocity $V$
right at the mean-field point.
For generic values of the coupling constant ($g\ne1$),
the mean jamming time $\mean{T}$ is an {\it increasing} function of $\xii$,
at least for a large enough system.
As shown by~(\ref{tlin}),~(\ref{tact}), and~(\ref{fss}),
it grows progressively as $\xii$ increases in general:
linearly in $N$ in the ballistic phase,
quadratically in $N$ right at the critical point ($\xii=\xiic$),
and exponentially in $N$ in the activated phase.
Right at the mean-field coupling ($g=1$) however,
the divergence law~(\ref{vc}) of the velocity implies that
$\mean{T}$ is a {\it decreasing} function of $\xii$.
Far from being paradoxical, such behaviour is only to be expected; in general
(far away from the mean-field coupling),
an increase of $\xii$ implies an increase of
locally felt frustration, and jamming times are thereby increased.
Right at the mean-field point however, there is in effect {\it no} local
frustration,
simply because the grains in the system are effectively non-interacting.
In its immediate vicinity,
an increased correlation length $\xii$ implies that clusters of interacting
grains of typical length $\xii$ reorganise themselves collectively to reach
a given ground state,
so that the jamming time decreases with increasing $\xii$.

\begin{figure}[!ht]
\begin{center}
\includegraphics[angle=90,width=.65\linewidth]{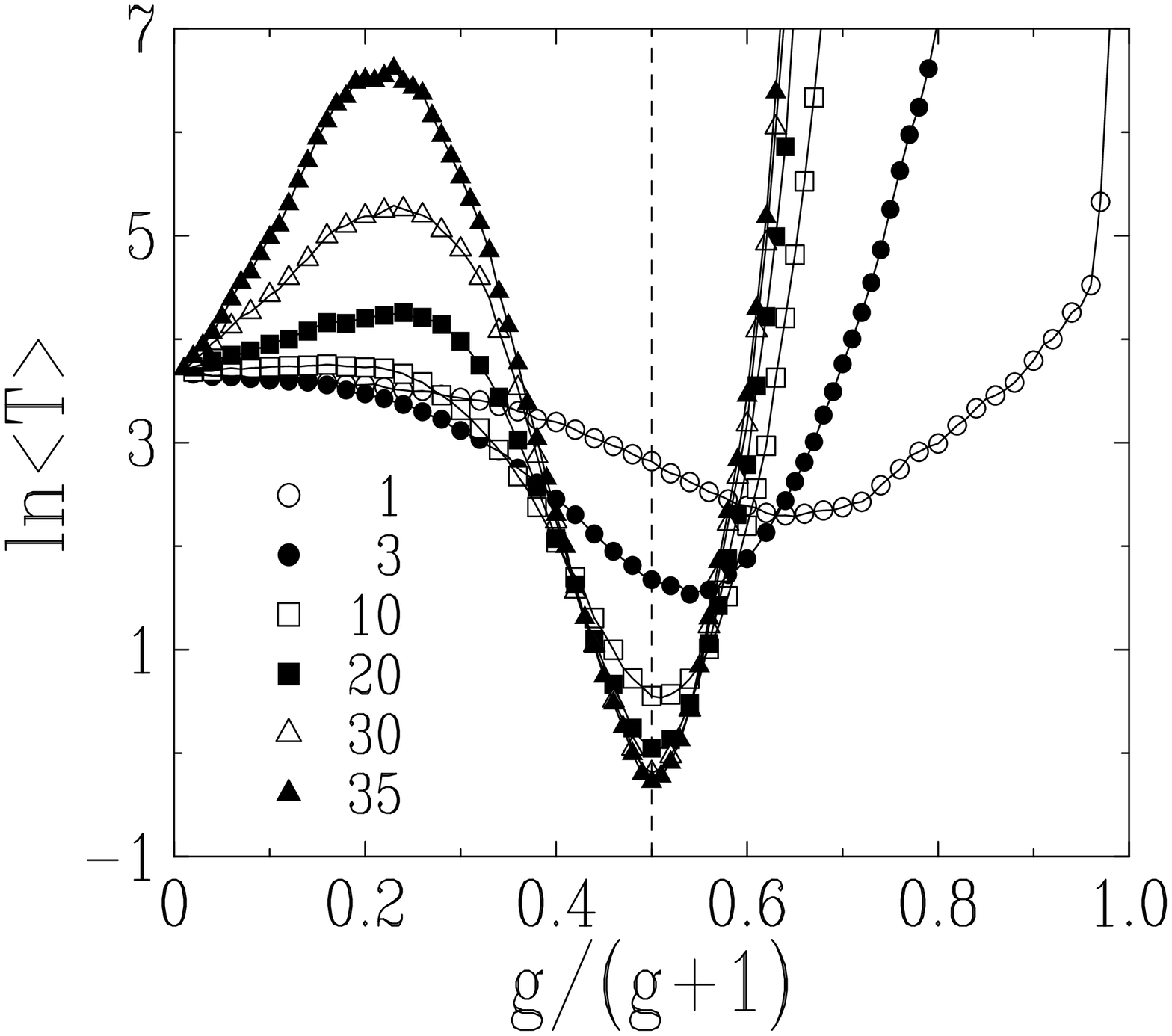}

\includegraphics[angle=90,width=.65\linewidth]{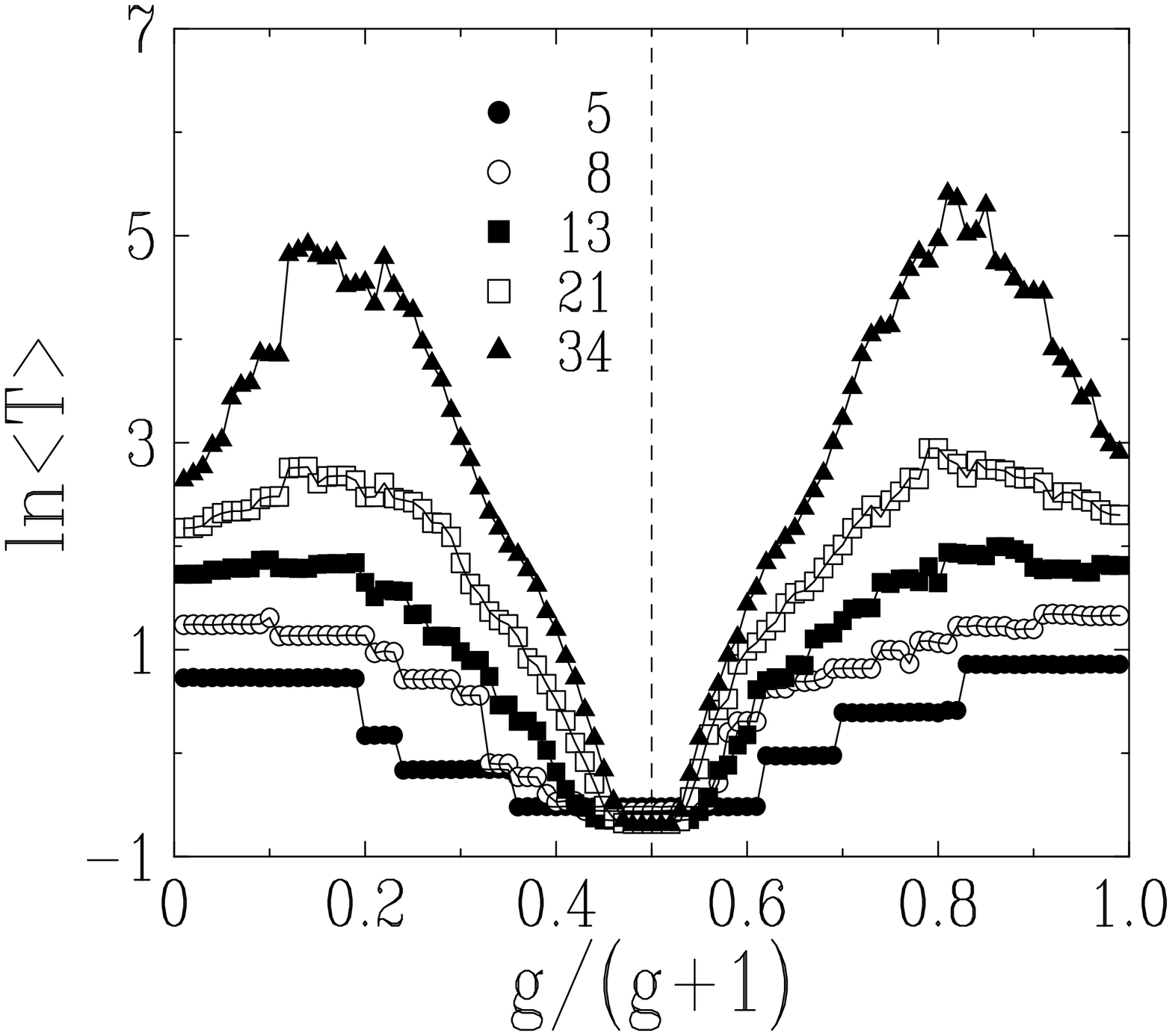}
\caption{\label{figtime}
Logarithmic plots of the mean jamming time $\mean{T}$ against $g/(g+1)$
for $\eps=1/\Phi$.
Top: $N=100$ and several values of $\xii$.
Bottom: $\xii=\infty$ and column sizes equal to Fibonacci numbers
from $F_5=5$ to $F_9=34$.
Dashed lines: mean-field coupling ($g=1$).}
\end{center}
\end{figure}

Figure~\ref{figtime} shows logarithmic plots of $\mean{T}$ against $g/(g+1)$
for $\eps=1/\Phi$.
The data on the upper panel correspond to $N=100$ and several values of $\xii$.
The jamming time is observed to have a highly non-monotonic dependence on~$g$.
All the data start from the same value ($\mean{T}\approx39$)
in the $g\to0$ limit.
For $\xii=1$ the data show a broad and shallow minimum.
As $\xii$ is increased, the data develop
a clear minimum near the mean-field coupling ($g=1$)
and a maximum around $g\approx 0.3$.
The maximum rises suddenly for $\xii\approx35$,
i.e., near the minimum value $\xiic\approx36$ of the critical line,
which is precisely reached for $g\approx 0.3$.
The observed rise therefore corresponds to the crossover between
the ballistic and activated phases.
The data on the lower panel correspond to $\xii=\infty$,
so that the system is in its extreme activated regime,
except in the immediate neighborhood of the mean-field coupling $g=1$.
The observed irregularities are genuine,
rather than being an artifact due to numerical noise;
it was noticed that,
if the column sizes were chosen to be successive Fibonacci numbers,
the amount of irregularity could be kept to a minimum.

\begin{figure}[!ht]
\begin{center}
\includegraphics[angle=90,width=.65\linewidth]{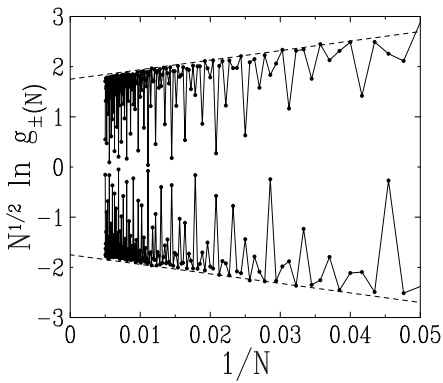}
\caption{\label{figdip}
Plot of the product $N^{1/2}\ln g_\pm(N)$ against $1/N$,
where $g_\pm(N)$ are the two coupling constants such that $\mean{T}=3$
for $\eps=1/\Phi$ and $\xii=\infty$.
Upper data: $g_+(N)>1$.
Lower data: $g_-(N)<1$.
Dashed lines have intercepts $\pm1.75$ and slopes $\pm19$.}
\end{center}
\end{figure}

The data on the lower panel of Figure~\ref{figtime}
show that the jamming time exhibits a dip around the mean-field coupling
($g=1$) in the $\xii=\infty$ limit,
which gets more and more symmetric, deep and narrow as $N$ is increased.
The width of this dip can be measured by
introducing two coupling constants $g_-(N)<1<g_+(N)$,
one on either side of the mean-field point, such that $\mean{T}=3$
(this value of the jamming time is chosen for convenience).
Figure~\ref{figdip} shows a plot of the products $N^{1/2}$ $\ln g_\pm(N)$
against $1/N$.
Both sequences of coupling constants are observed to behave very
symmetrically, and to depend on $N$ in a very irregular way.
The hulls of the data however converge to the non-trivial limits $\pm1.75$,
implying the scaling law
\beq
\delta g_+(N)\approx -\delta g_-(N)\sim N^{-1/2}.
\label{dipfss}
\eeq
We have indeed $\delta g=g-1\approx\ln g$ for $\delta g\ll1$, in conformity
with the ordinate of Figure~\ref{figdip}.

The reason for the above scaling behaviour can be explained as follows.
For $\xii=\infty$, the local field $H_n$ acting on grain $n$ reads
\beq
H_n=\eta-f(\s_n)+\delta g\sum_{m=n+1}^Nf(\s_m),
\label{mfp}
\eeq
where $\eta$ is the mean field introduced in~(\ref{etadef}).
In a random configuration of grain orientations,
the difference $H_n-\eta$ can thus be evaluated to be of order $(\delta
g)N^{1/2}$, by the law of large numbers.
This estimate has two consequences.
First, requiring that the difference $H_n-\eta$ is of order unity
allows one to recover~(\ref{dipfss}).
Second, the above estimate can also be viewed as a finite-size scaling form
of the result~(\ref{xone}).
The width $\delta g\sim N^{-1/2}$ of the dip
is indeed expected to be such that $\xii$ is comparable to $N$.
This yields the quadratic divergence $\xiic\sim1/(\delta g)^2$, as
in~(\ref{xone}).

Finally, the irregularities visible in the bottom panel
of Figure~\ref{figtime}, and especially in Figure~\ref{figdip},
suggest an analogy with the phenomenon of defect nucleation,
encountered in earlier work~\cite{IV} to explain intermittency
in the case of a weakly tapped ($\T\ll1$) column of irregular grains.
Around the mean-field point, the predominant behaviour is of course
dominated by the mean field; however, occasional perturbations
in the form of excess $j_n$ and $h_n$ fields can lead to the nucleation
of defects at specific sites $n$,
generating the self-similar patterns observed.

\subsection*{R\^ole of a finite $\xidy$}

So far we have limited the discussion of the dynamical phase diagram
to the case where activation energies are negligible, so that $\xidy=\infty$.
The r\^ole of a finite dynamical length $\xidy$ has already been investigated
in the $g\to0$ regime for $\eps=1$~\cite{V}.
Its main effect is to induce two novel dynamical phases,
whose main characteristics are as follows.

\smallskip\noindent$\bullet$ {\it Logarithmic phase.}
The logarithmic phase replaces the ballistic one for $\xii<\xiic$
when $\xidy$ is much smaller than $N$.
In this regime, the system still orders from above,
but the growth of the thickness $L(t)$ of the upper ordered layer
is slowed down by gravity, according to the local frequencies~(\ref{odef}).
We have therefore
\beq
\frac{\dy L}{\dy t}\approx V\,\exp\left(-\frac{L}{\xidy}\right).
\eeq
The results~(\ref{vt}) and~(\ref{tlin}) are recovered for $\xidy\gg N$,
i.e., in the ballistic phase.
In the logarithmic phase, when $\xidy\ll N$,
the thickness of the ordered layer is predicted to grow logarithmically:
\beq
L(t)\approx\xidy\ln\frac{Vt}{\xidy},
\eeq
so that the jamming time diverges exponentially fast:
\beq
\mean{T}\approx\frac{\xidy}{V}\,\exp\left(\frac{N}{\xidy}\right).
\label{tlog}
\eeq

\smallskip\noindent$\bullet$ {\it Glassy phase.}
The glassy phase replaces the activated one when $\xidy$ is sufficiently small.
The crossover between the activated and glassy phases
takes place when the exponential growth of the slowest local time scale
of the column, $1/\o_N=\e^{N/\xidy}$
(which also governs the jamming time~(\ref{tlog}) in the logarithmic phase),
exceeds the entropic growth
$\mean{T}\sim\e^{aN}$ characteristic of the activated phase.
This line of reasoning predicts that the glassy phase
can only be observed if $\xidy$ is a microscopic length: $\xidy<1/a$.
The jamming time diverges exponentially with $N$
in the glassy phase, according to $\mean{T}\sim\e^{N/\xidy}$.

Since the dependence on $\xidy$ remains unchanged with respect
to the earlier case (see~\cite{V}), the purely depth-dependent features of the
glassy and logarithmic phases remain the same.
Among the novel features of the present model,
we mention one related to the existence of a mean-field point.

\begin{figure}[!ht]
\begin{center}
\includegraphics[angle=90,width=.65\linewidth]{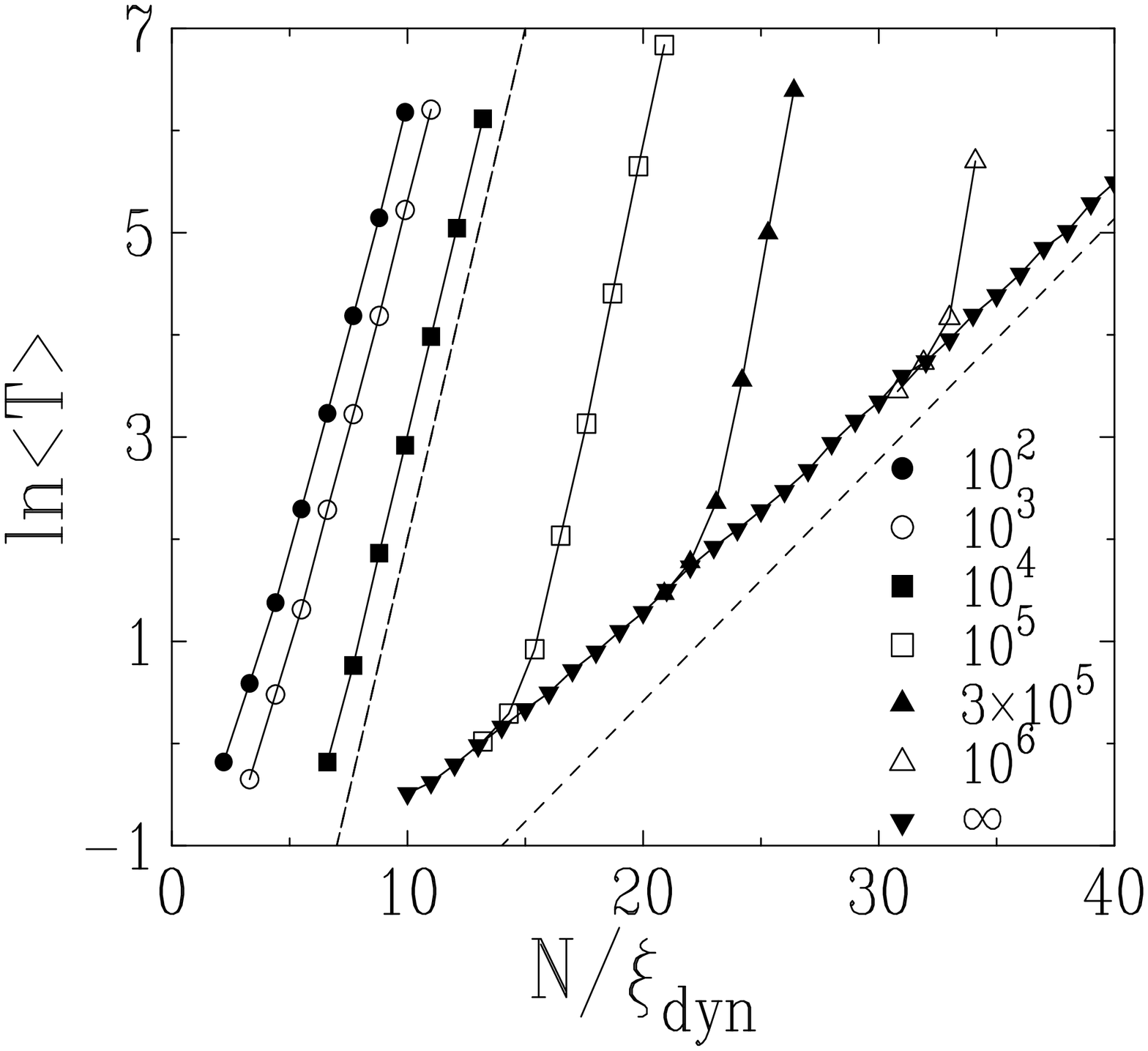}
\caption{\label{figtimexi}
Logarithmic plot of the mean jamming time $\mean{T}$
against the ratio $N/\xidy$, for $\eps=1/\Phi$, $g=1$, $\xidy=50$,
and several values of $\xii$.
The left dashed line has the unit slope corresponding to~(\ref{tlog}).
The right one has the slope corresponding to~(\ref{tmf}),
i.e., $(1-\eps)/(1+\eps)=1/\Phi^3\approx0.236067$.}
\end{center}
\end{figure}

Consider the model right at the mean-field coupling ($g=1$).
In the case of mean-field statics and slow dynamics,
i.e., for $\xii=\infty$ and $N\gg\xidy$,
the jamming time grows exponentially with $N$,
with an $\eps$-dependent prefactor given by~(\ref{tmf}).
In the generic situation where the column size $N$
is much larger than both lengths $\xidy$ and $\xii$,
the jamming time also grows exponentially with $N$,
albeit with the `trivial' prefactor of~(\ref{tlog}).
Figure~\ref{figtimexi} presents a logarithmic plot of $\mean{T}$
against the ratio $N/\xidy$, for $\eps=1/\Phi$, $g=1$, $\xidy=50$,
and several values of~$\xii$.
Both limiting growth rates are clearly observed,
the crossover between both regimes occurring
for very large values of~$\xii$.
We note again that the jamming time {\it decreases} with increasing $\xii$ in
the vicinity of the mean-field limit, consistent with the picture presented
earlier.
Finally, in order to avoid having large irregularities,
$N$ has been chosen to be a multiple of the Fibonacci number $F_{10}=55$.

\section{Dynamics: attractors}
\setcounter{equation}{0}
\def\theequation{5.\arabic{equation}}

We next turn to the statistics of attractors sampled by zero-temperature dynamics,
starting from a random initial configuration.
This question is interesting because
of its relationship with Edwards' flatness hypothesis~\cite{edwards};
according to this, attractors are sampled uniformly,
so that the static entropy $\S$ (see~(\ref{ssdef}))
and the dynamical entropy~$S$ (see~(\ref{sddef})) coincide.

For the time being, we consider the case where $\xidy=\infty$.
Attractor statistics have been investigated in~\cite{V}
for $\eps=1$ in the $g\to0$ limit,
where ground states have a simple characterisation:
they are all the dimerised configurations.
In the present case, for generic values of the parameters,
the problem is more difficult because ground states are not known a priori,
and most certainly do not allow for a simple static characterisation.
In this respect the present situation is similar to that
of tapping dynamics on various models,
e.g.~the Kob-Andersen model~\cite{tapping}.
In the following we present data illustrating what are, according to us,
the main features of the statistics of attractors.

First, in order to quantify the r\^ole of the coupling constant~$g$,
we have chosen to focus on the dynamical overlap
between attractors $\s_n^{(g)}$ at coupling $g$
and $\s_n^{(0)}$ at infinitesimal coupling ($g=0^+$), defined as
\beq
\Omega=\frac{1}{N}\sum_{n=1}^N\mean{\s_n^{(g)}\s_n^{(0^+)}}.
\eeq
Note that here the system is started in {\it the same} initial configuration
and subjected to {\it the same} stochastic noise
(i.e., in practice, the same sequence of random numbers).

\begin{figure}[!ht]
\begin{center}
\includegraphics[angle=90,width=.65\linewidth]{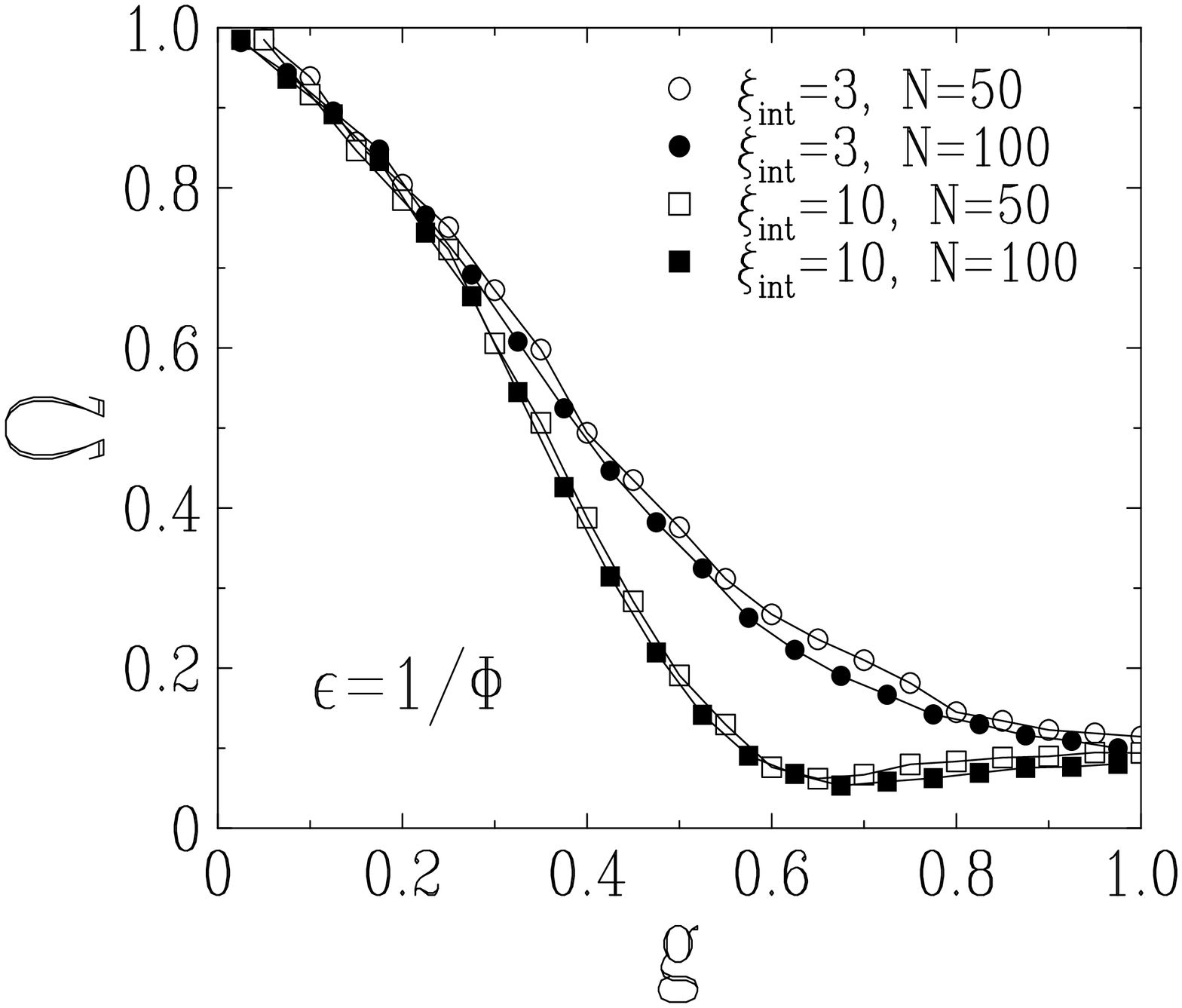}

\includegraphics[angle=90,width=.65\linewidth]{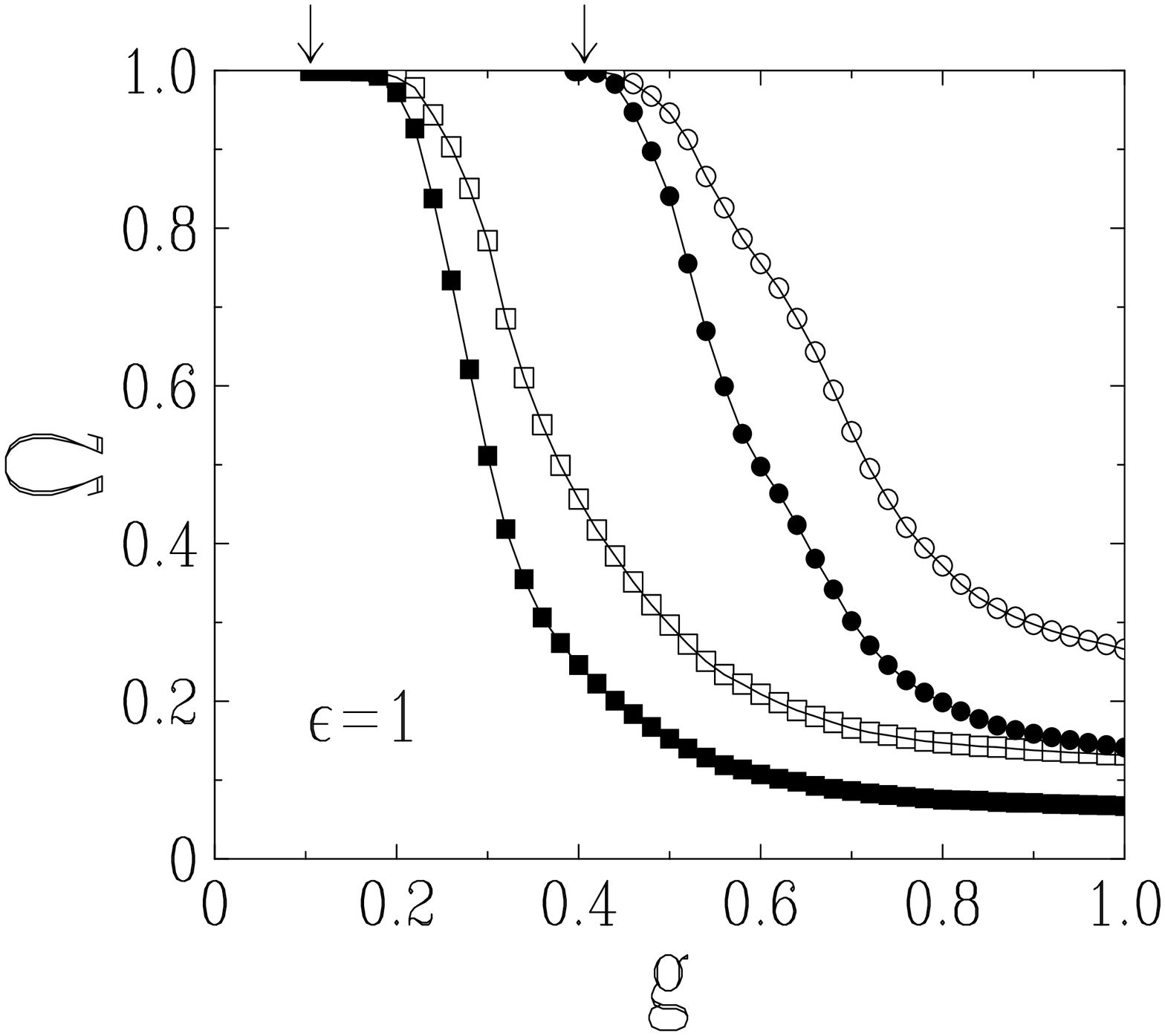}
\caption{\label{figoverlap}
Plot of the dynamical overlap $\Omega$ against $g$
for both shape parameters $\eps=1/\Phi$ (top) and $\eps=1$ (bottom)
and the same values of $\xii$ and $N$.
Arrows on top of the lower panel:
dynamical threshold $g_\dy\approx0.395612$ for $\xii=3$ and
$g_\dy\approx0.105170$ for $\xii=10$.}
\end{center}
\end{figure}

Figure~\ref{figoverlap} shows data for $\Omega$
for $\eps=1$ and $\eps=1/\Phi$, $N=50$ and 100,
and two values of the interaction length ($\xii=3$ and 10)
deep in the ballistic phase, i.e., much smaller than $\xiic$.
The overlap behaves differently in both cases.
For the irrational shape parameter $\eps=1/\Phi$ (top),
the dynamics at infinitesimal coupling
drive the system to its unique quasiperiodic ground state,
so that~$\Omega$ is nothing but the overlap
between the finite-coupling attractor $\s_n^{(g)}$ and that ground state.
This overlap exhibits a continuous decay as a function of $g$,
which is remarkably size-independent.
This suggests the following picture: quasiperiodic ordering spreads from the
top of the column in the ballistic regime, as does the splitting of
the quasiperiodic state into finite patches of length $\L(g)\sim 1/g$ at finite
coupling.
For a given value of $g$, the orientation of a grain is fixed, depending on
its position in one such patch.
Increasing the length of the column leaves this orientation unchanged, only
adding on more, similar patches corresponding to a given attractor, and thus
leaving the overlap function unchanged.
For the rational shape parameter $\eps=1$ (lower panel of
Figure~\ref{figoverlap}),
the overlap behaves in a very different manner.
It remains equal to unity in the range $0<g<g_\dy$,
where the values of the dynamical threshold $g_\dy$ (see~(\ref{gd}))
are shown as arrows.
This is expected, as the dynamics are strictly independent of $g$ in
that range.
The overlap then falls off very abruptly, more and more so for larger systems.

We now turn to the statistics of attractors {\it per se}.
Let us first consider in detail the case $\eps=1$.
In the weak-coupling phase ($0<g<g_\st$),
attractors are known to consist of dimers.
It is thus natural to characterise attractors by their contents in (polarised)
dimers all over the phase diagram.
The main difference between the present general situation and that of the
weak-coupling regime is that now attractors are only {\it partially} dimerised.
We are thus led to introduce {\it two} local dimer order parameters:
\beq
\Delta_k=-\mean{\s_{2k-1}\s_{2k}},\quad
\Pi_k=\frac{1}{2}\mean{\s_{2k}-\s_{2k-1}}.
\label{deltaloc}
\eeq
The first order parameter is such that $\Delta_k=1$
whenever there is a dimer at the $k$-th position and
irrespective of its polarisation, whereas $\Delta_k=-1$ otherwise.
The second one is sensitive to the polarisation of the dimer at the $k$-th
position; it is such that $\Pi_k=+1$ if there is a $-+$ dimer,
$\Pi_k=-1$ if there is a $+-$ dimer, and $\Pi_k=0$ otherwise.
Both local order parameters vanish on average in a random configuration,
where orientations are uncorrelated.
Assuming that the column consists of an even number of grains ($N=2K$),
we also define global order parameters as spatial averages of the local ones:
\beq
\Delta=\frac{1}{K}\sum_{k=1}^K\Delta_k,\quad
\Pi=\frac{1}{K}\sum_{k=1}^K\Pi_k.
\label{deltaglo}
\eeq
The polarisation-sensitive order parameter $\Pi_k$ can be recast in terms
of the staggered orientation profile
\beq
S_n=(-1)^n\mean{\s_n}
\label{stagdef}
\eeq
as
\beq
\Pi_k=\frac{1}{2}(S_{2k}+S_{2k-1}),
\eeq
so that
\beq
\Pi=\frac{1}{N}\sum_{n=1}^NS_n
\eeq
is nothing but the mean staggered orientation.

\begin{figure}[!ht]
\begin{center}
\includegraphics[angle=90,width=.65\linewidth]{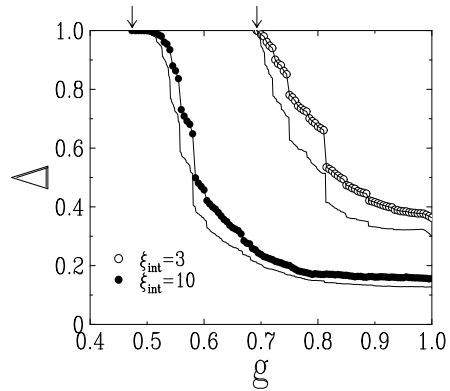}
\caption{\label{figdimer}
Plot of the global dimer order parameter $\Delta$ of attractors against $g$
for $\eps=1$, $N=20$, and two values of $\xii$.
Lines with symbols: dynamical ensemble.
Lines without symbols: static ensemble.
Arrows on top: static threshold $g_\st\approx0.692394$ for $\xii=3$ and
$g_\st\approx0.473376$ for $\xii=10$.}
\end{center}
\end{figure}

Figure~\ref{figdimer} shows a plot of the global dimer order parameter
$\Delta$ of the attractors against $g$, for $\eps=1$, $N=20$, and $\xii=3$ and~10.
This modest system size has been chosen in order to test Edwards'
flatness hypothesis fully.
The plot indeed presents a comparison between
(i) the static (or a priori) ensemble, where all attractors
are obtained by means of an exact enumeration and taken with equal weights,
and (ii) the dynamical ensemble, where attractors are sampled
according to the dynamics, with a random initial configuration.
Data pertaining to both ensembles behave similarly.
They remain equal to unity in the whole range $0<g<g_\st$,
where the values of the static threshold~$g_\st$ (see~(\ref{gs})) are shown as
arrows.
Both datasets then fall off in a similar way
(including their fine structure due to the finite size of the column),
thus indicating that attractors are sampled rather uniformly by the dynamics.
In other words, Edwards' hypothesis, although not exact,
provides a good approximation to the attractor statistics in this ballistic regime.

Next, in order to investigate the possible consequences
of slow (activated) dynamics on the statistics of attractors,
we will deal with larger values of $\xii$ and with arbitrary values of $\eps$.
A natural observable in this case~\cite{V} is the pseudo-energy $\E$ per grain,
defined as
\beq
\E=-\frac{1}{N}\sum_{n=1}^NH_n\s_n.
\label{edef}
\eeq
This definition can be motivated as follows.
If the $\s_n$ were independent spins in external fields $H_n$,~(\ref{edef})
would be the corresponding Hamiltonian.
In the present model, the local fields $H_n$ depend on the orientations $\s_m$
in a non-symmetric way, so that the dynamics do not obey detailed balance,
and the statics are not described by a Hamiltonian.
The pseudo-energy defined by~(\ref{edef}) however
provides a useful measure of the amount of disorder,
either in an arbitrary configuration or in an attractor.

\begin{figure}[!ht]
\begin{center}
\includegraphics[angle=90,width=.65\linewidth]{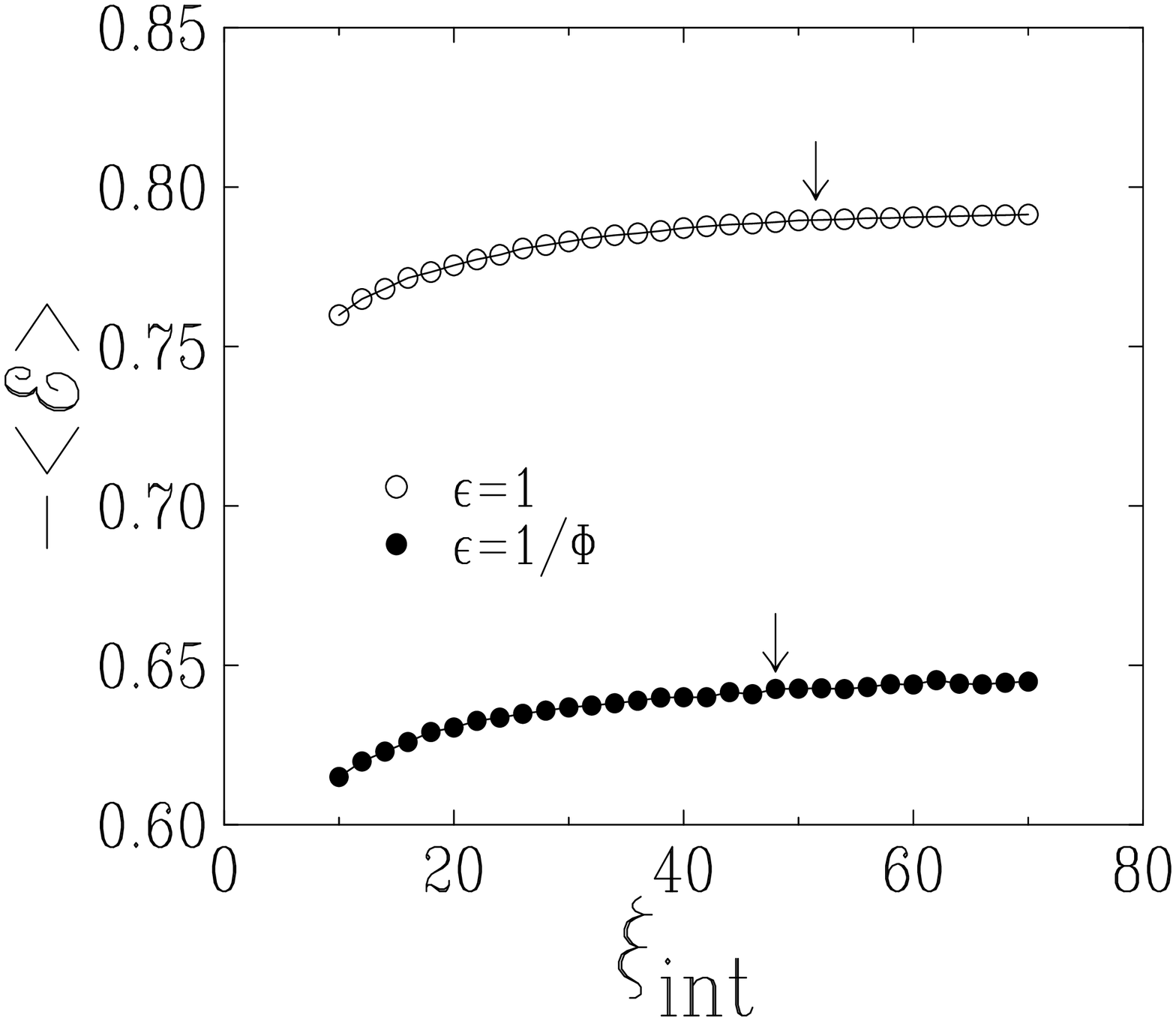}

\includegraphics[angle=90,width=.65\linewidth]{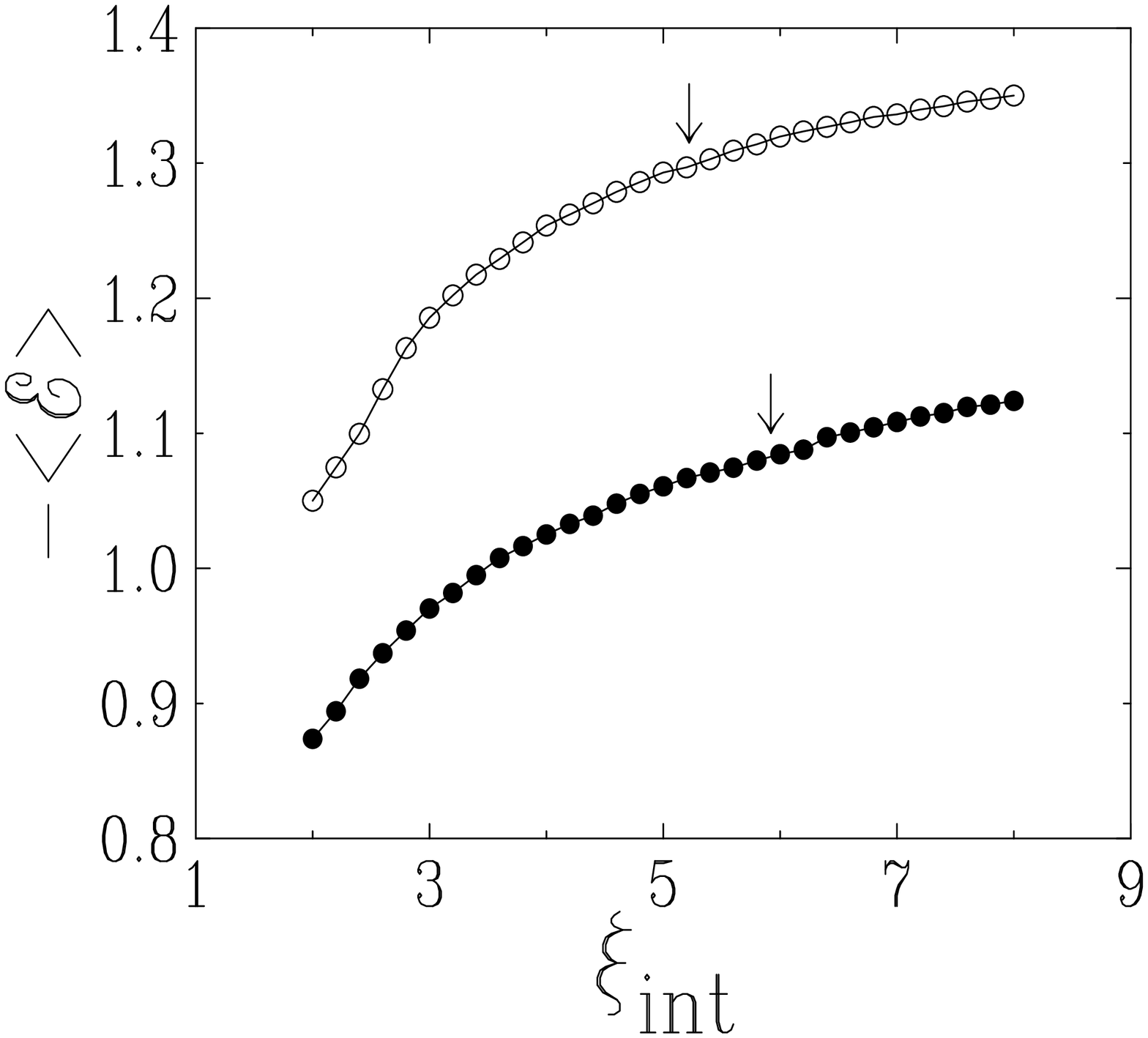}
\caption{\label{fige}
Plot of (minus) the mean attractor pseudo-energy $\mean{\E}$ per grain
against $\xii$, for $N=200$ and both shape parameters.
Arrows: critical values $\xiic$.
Top: $g=0.6$.
Bottom: $g=2$.}
\end{center}
\end{figure}

We have chosen to focus on two values of the coupling constant,
one on each side of the mean-field coupling, i.e., $g=0.6$ and $g=2$,
where the critical value of the interaction length is respectively large and small.
These are $\xiic\approx51$ for $\eps=1$ and 48 for $\eps=1/\Phi$ when $g=0.6$;
and $\xiic\approx5.2$ for $\eps=1$ and 5.9 for $\eps=1/\Phi$ when $g=2$.
Figure~\ref{fige} shows plots of (minus) the mean pseudo-energy~$\mean{\E}$ per grain
against $\xii$ in a range containing the critical values $\xiic$
(shown by arrows) which separate the ballistic and the activated phases.
The mean energy exhibits a very weak and regular increase as a function
of the interaction length $\xii$; this verifies what we might expect, that
increasing correlations will increase the amount of order in the system.
The mean energy however, shows no anomalies at all as the critical point is
crossed;
this smooth behaviour is to be contrasted with the explosive rise in
jamming times which accompanies it.
Over the ranges of values of~$\xii$ corresponding to the plotted data,
the mean jamming time indeed increases by factors of 150 for $\eps=1$ and 580
for $\eps=1/\Phi$ in the case of $g=0.6$, and by factors of
2100 for $\eps=1$ and 80 for $\eps=1/\Phi$ in the case of $g=2$.

\subsection*{R\^ole of a finite $\xidy$}

The effect of a finite dynamical length $\xidy$
on the statistics of attractors has already been investigated in~\cite{V}
for $\eps=1$ and $g\to0$, that is, for weak frustration.
The main qualitative conclusion there was that a truly non-trivial sampling of
attractors was observed only in the glassy phase,
whereas the attractor statistics in the three other dynamical phases
were found to be in qualitative agreement with Edwards' flatness hypothesis.
The effect of increasing frustration, in the present paper, might be expected
only to enhance the non-triviality in the sampling of attractors in the glassy
phase, thus deepening the contrast between this and the other three phases.

\begin{figure}[!ht]
\begin{center}
\includegraphics[angle=90,width=.65\linewidth]{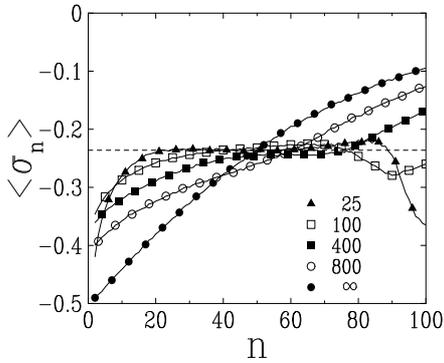}
\caption{\label{figproxi}
Plot of the orientation profile of the attractors again depth $n$
for $N=100$, $\eps=1/\Phi$, at the mean-field coupling $g=1$, and $\xidy=50$.
Symbols: data for several values of $\xii$.
Dashed line: mean orientation $\mean{\s}=-1/\Phi^3\approx-0.236067$.}
\end{center}
\end{figure}

We consider first the vicinity of the mean-field point.
Figure~\ref{figproxi} shows the orientation profile $\mean{\s_n}$ of the
attractors for parameters similar to those used in Figure~\ref{figtimexi},
i.e., $\eps=1/\Phi$, $g=1$, $N=100$, and $\xidy=50$.
In the case of mean-field statics ($\xii=\infty$),
a non-trivial orientation profile is observed,
similar to that shown in Figure~\ref{figpro}.
In the generic situation where the interaction length $\xii$ is finite,
the profile soon becomes very nearly flat,
and therefore equal to its mean value~(\ref{aves}), i.e.,
$\mean{\s}=-1/\Phi^3\approx-0.236067$,
thus indicating that attractors are sampled nearly uniformly by the dynamics.

Next, we explore the statistics of attractors for a generic coupling $g$,
to, in particular, test the validity of Edwards' hypothesis.
We use a coupling constant greater than the mean field value ($g>1$),
so as to be able to have a relatively small $\xiic$
with our choice of boundary condition~(\ref{upper}).
The consequence of this is that the slow phases (activated and glassy)
are easily observed with a modest system size.
This possibility of tuning $\xiic$, and consequently system sizes,
to reasonable values was not present in the case of the
weak-coupling regime~\cite{V}, where lengths were perforce large;
it explains in part our rationale for not changing
the boundary condition~(\ref{upper}),
of which more will be said in the concluding section.

\begin{figure}[!ht]
\begin{center}
\includegraphics[angle=90,width=.65\linewidth]{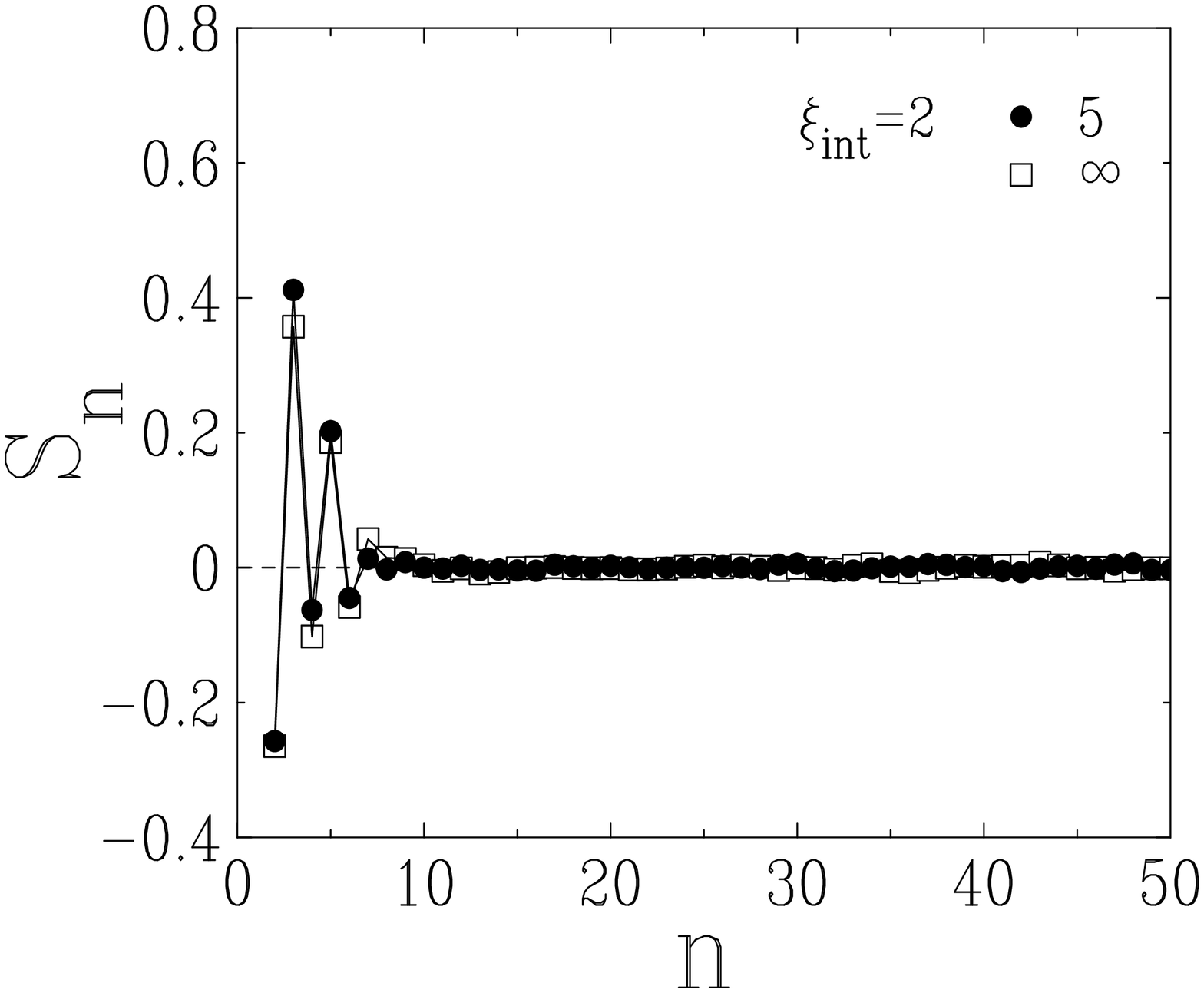}

\includegraphics[angle=90,width=.65\linewidth]{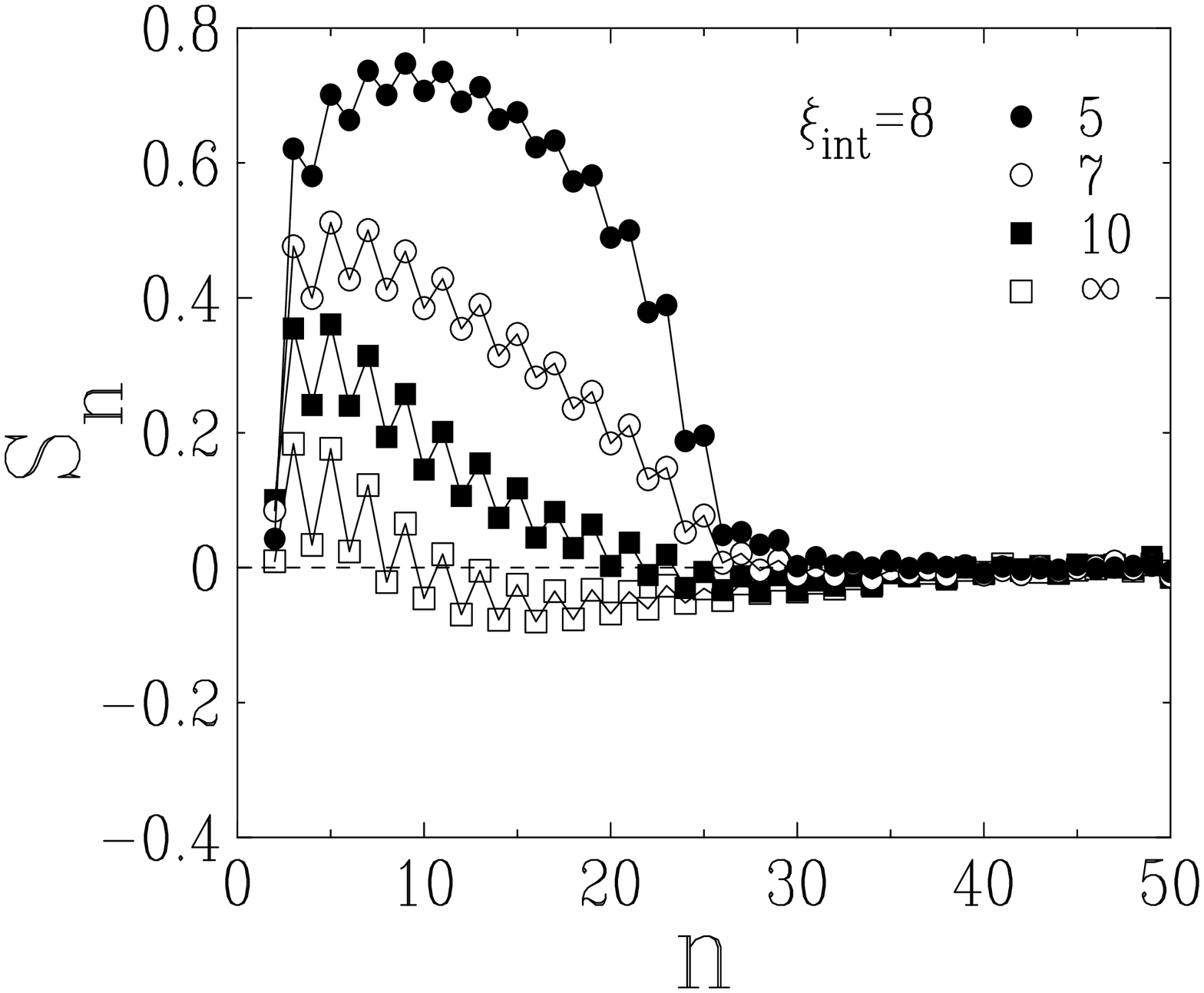}
\caption{\label{figs28}
Plot of the staggered orientation profile of the attractors
against depth $n$ for $\eps=1$, $g=2$, $N=50$,
and several values of $\xidy$.
Top: $\xii=2$.
Bottom: $\xii=8$.}
\end{center}
\end{figure}

\begin{figure}[!ht]
\begin{center}
\includegraphics[angle=90,width=.65\linewidth]{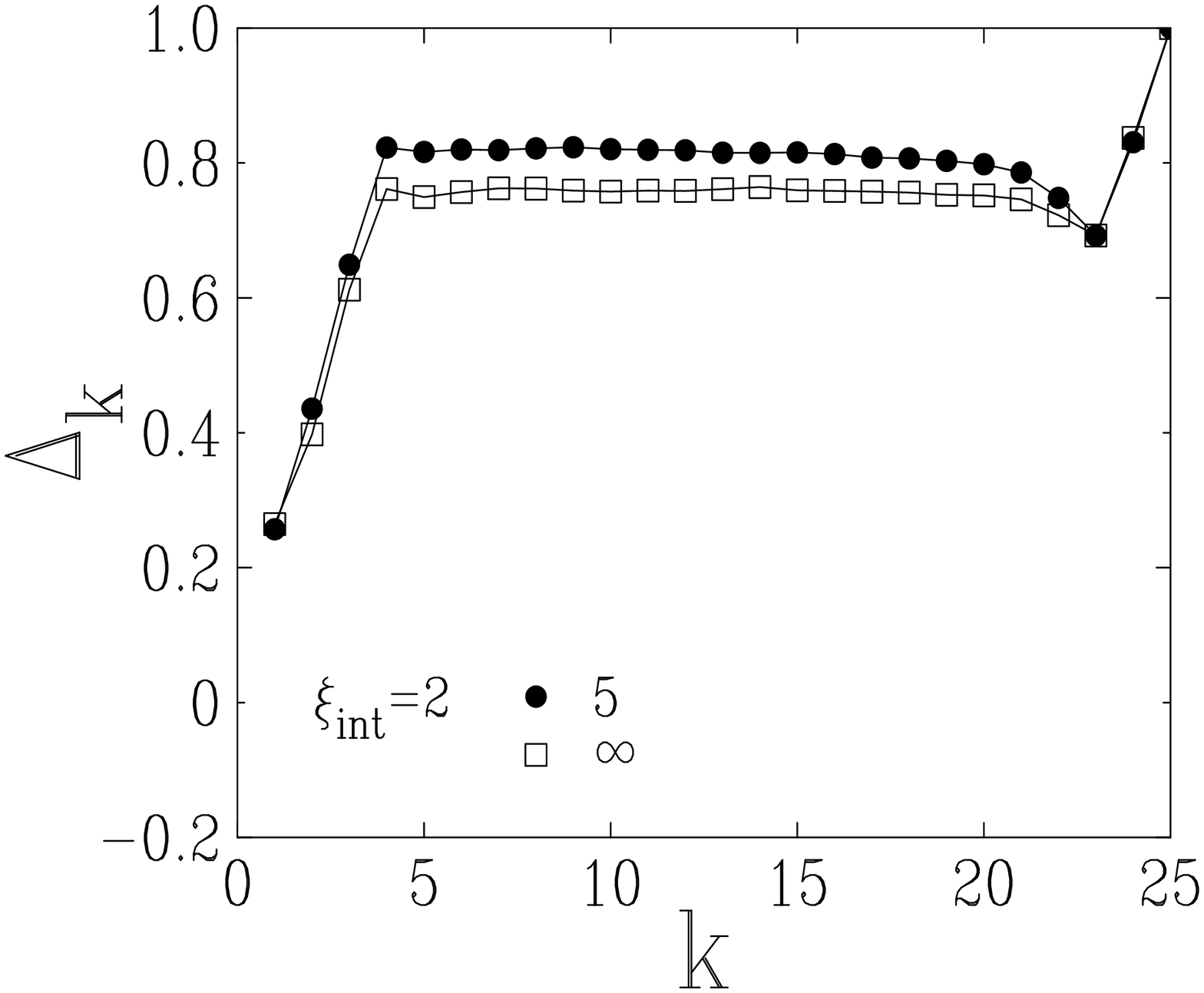}

\includegraphics[angle=90,width=.65\linewidth]{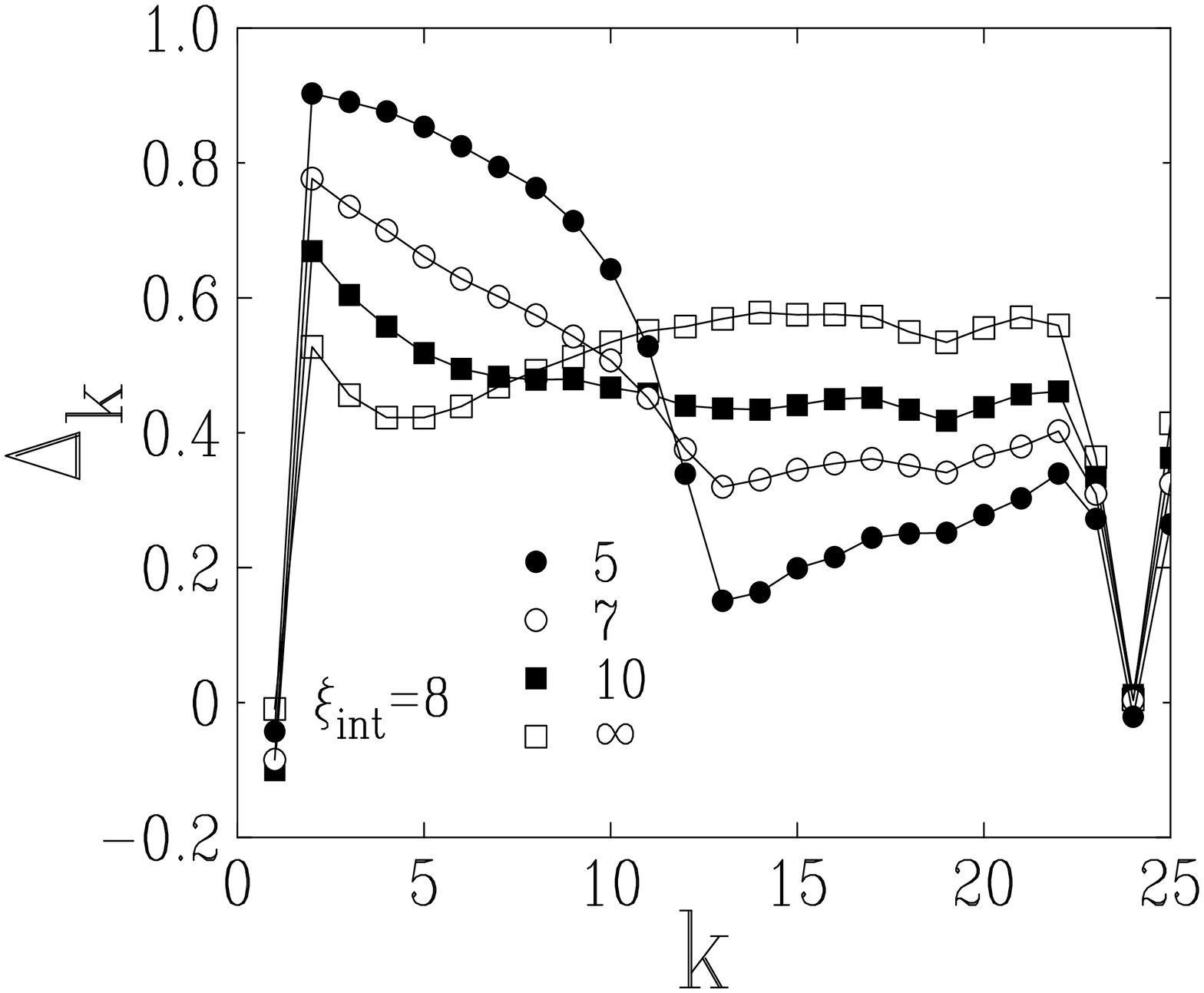}
\caption{\label{figd28}
Same as Figure~\ref{figs28} for the dimer order parameter~$\Delta_k$,
plotted against the dimer number $k$.}
\end{center}
\end{figure}

\begin{figure}[!ht]
\begin{center}
\includegraphics[angle=90,width=.65\linewidth]{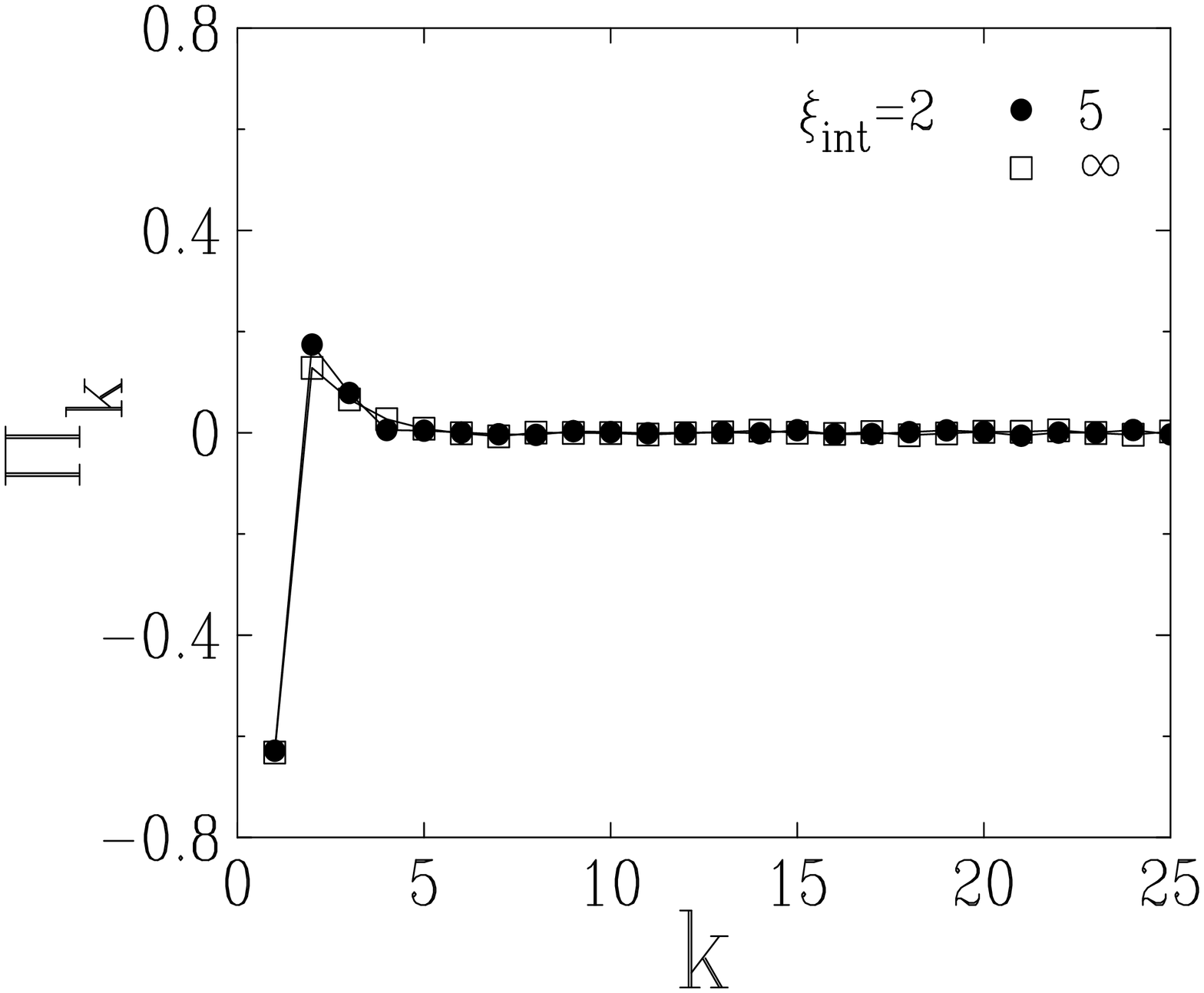}

\includegraphics[angle=90,width=.65\linewidth]{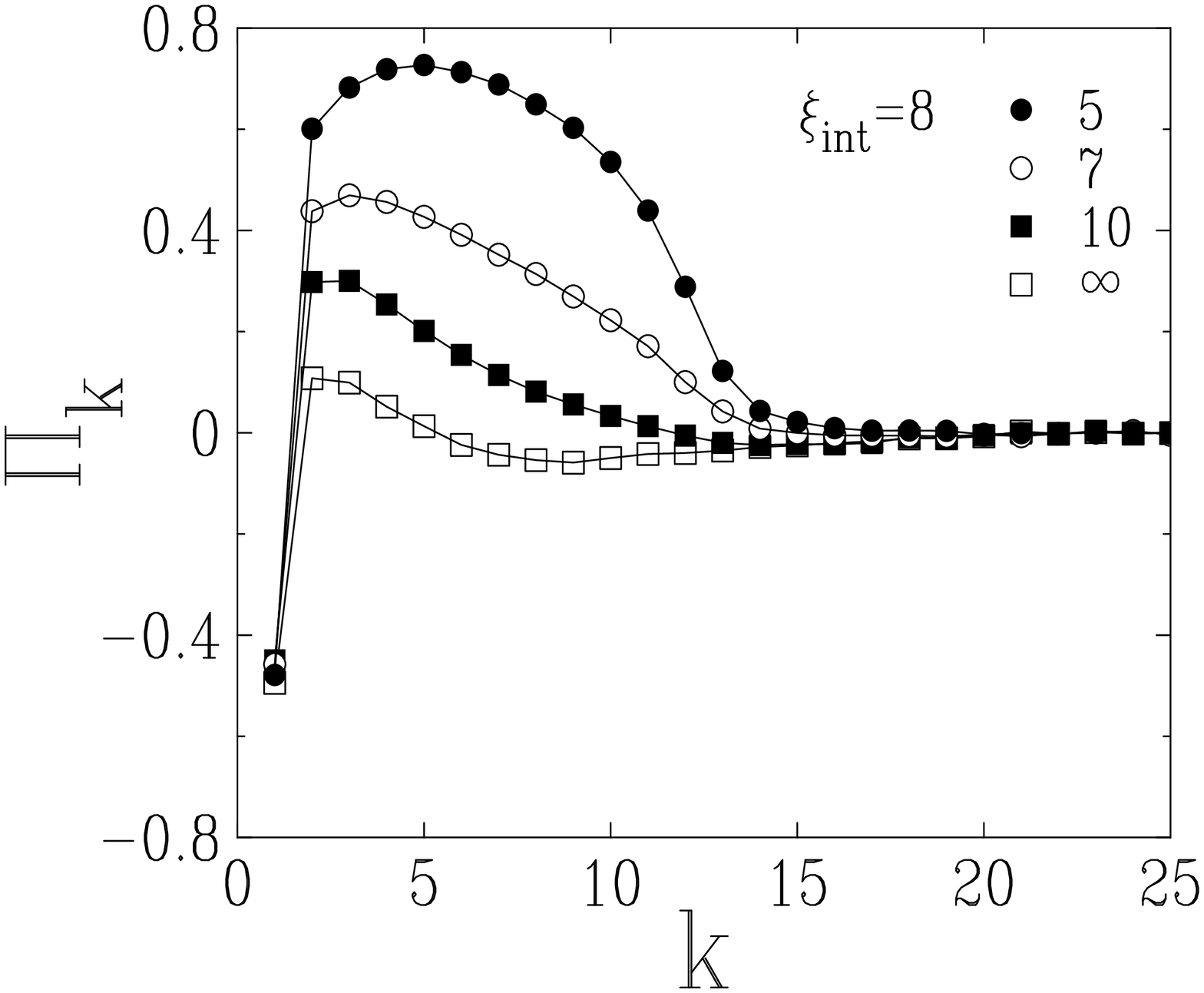}
\caption{\label{figp28}
Same as Figure~\ref{figs28} for the polarisation-sensitive dimer order
parameter $\Pi_k$, plotted against the dimer number $k$.}
\end{center}
\end{figure}

We choose $\eps=1$,
in order to keep using the local dimer order parameters~(\ref{deltaloc}),
a coupling constant of $g=2$, so that $\xiic\approx5.2$,
and a system size $N=50$.
We compare the nature of the attractors
at $\xii=2$ (below $\xiic$: ballistic to logarithmic crossover)
and $\xii=8$ (above $\xiic$: activated to glassy crossover).
Figures~\ref{figs28} to~\ref{figp28}
respectively show plots of the staggered orientation profile $S_n$
and of the local dimer order parameter $\Delta_k$ and $\Pi_k$
for several values of $\xidy$ spanning the crossovers.

Our earlier speculations are concretised by the following observations.
The data for $\xii=2$,
shown in the upper panels of Figures~\ref{figs28} to~\ref{figp28},
and pertaining to the crossover between the ballistic and the logarithmic
phases, hardly show any dependence on $\xidy$,
i.e., on the fast (ballistic) or slow (logarithmic) nature of the dynamics.
Since we have established that Edwards' flatness holds,
at least qualitatively, for the ballistic phase,
this is a clear indication that it holds also for the logarithmic phase.
In terms of the actual ordering, we
note the high and nearly uniform value of the dimer order parameter
($\Delta_k\approx0.75$) in the upper panel of Figure~\ref{figd28};
the picture is that the attractor is most ordered at the base, and more or
less ordered (dimerised) throughout its length, except near the very top.
The staggered orientation profile and the order parameter $\Pi_k$,
shown in the upper panels of Figures~\ref{figs28} and~\ref{figp28},
demonstrate that the dimers are essentially unpolarised,
except in a thin boundary layer near the top of the column.

On the other hand, the data for $\xii=8$,
shown in the lower panels of Figures~\ref{figs28} to~\ref{figp28},
and pertaining to the crossover between the activated and the glassy phases,
exhibit a strong dependence on $\xidy$.
Highly non-trivial profiles
are observed for the smaller values of $\xidy$, and especially for $\xidy=5$,
which can be considered as glassy, given the modest size of the column.
In this case, we see that the ensuing structure of the column
is highly heterogeneous.
The large values of the parameters $\Delta_k$ and $\Pi_k$
observed in the upper part of the column indicate a large degree of dimerisation,
most of the dimers being polarised as $-+$.
In other words, there is a definite preference for {\it one} of the
`crystalline' arrangements of dimers found
in earlier work~\cite{V} in the glassy regime for $\eps=1$ and $g\ll1$.
The choice of one preferred direction of polarisation can be explained as follows.
The boundary condition $\s_1=+1$, together with the observed smallness of
$\mean{\s_2}$, yields a definitely negative~$h_3$, and hence a trend toward
$\mean{\s_3}<0$, i.e., $\Pi_2>0$.
For $\xii=8$ and $\xidy=5$,
$\mean{\s_2}\approx0.04$ is indeed much smaller than $\mean{\s_3}\approx-0.62$.
The same polarising effect acts on the deeper dimers as~well.
Another notable difference with respect to the weak-coupling situation
considered in~\cite{V} is that the lower part of the column
appears much less ordered.
One may speculate that a disordered lower part is a generic characteristic
of attractors in the glassy regime.
Grains are indeed very slow as soon as $n\gg\xidy$, and thus hardly equilibrate.
The weak-coupling regime for $\eps=1$ appears as an exception to this general rule,
because there it is already known from statics
that all ground states are fully dimerised.

\section{Discussion}

The full phase diagram of the frustrated column model has been presented in
this work.
This model of a column of grains has been developed~\cite{II}
and investigated in earlier work, first in the directed situation ($g=0$)
for arbitrary values of the shape parameter $\eps$~\cite{III,IV},
and then in the weak-coupling regime of the symmetric case ($\eps=1$,
$g\ll1$)~\cite{V}.
The present work is the first investigation of the model over its entire
parameter space,
i.e., for generic values of $\eps$, $g$, and of both lengths $\xii$ and $\xidy$.
One of the most novel features is the existence of a mean-field point ($g=1$)
where the erstwhile local constraints on grain orientations disappear,
becoming global;
the model in this limit is similar to one of non-interacting grains,
which has been analysed in~\cite{II}.

The case of $\eps=1$ provides a useful illustration of many features of the
model, including the physical nature of the mean-field limit.
In this symmetric situation
there are essentially two ways in which the average orientation of zero can be
achieved:
by dimerised packings in the presence of strong local compacting constraints
($g\to0$),
and by uniformly random packings such that there are equal numbers of ordered
and disordered grains globally, in the absence of local constraints ($g=1$).
Most of the features of the phase diagram for $\eps=1$ for $g<1$
can be explained in terms of an interpolation between these two extremes.
Beyond the mean-field limit ($g>1$),
the picture is one of a column of grains that is strongly frustrated,
partly due to the chosen boundary condition (the uppermost spin is fixed),
so that the activated phase extends down to arbitrary small values of~$\xii$
at strong coupling.

This is a natural point at which to discuss the implications
of our choice ~(\ref{upper}) of boundary condition.
For the directed model~\cite{III,IV}, where order propagated downwards,
it was a natural choice to fix the top spin.
This choice may seem to become less and less natural as increasing $g$,
as the effects of the frustrating field $j_n$ made
the propagation of order less and less directional.
Indeed, it might be argued that, in the large-coupling regime,
the natural choice would be to fix the bottom spin,
corresponding to the predominant upward propagation of order.
There are two aspects to this issue.
On the one hand,
fixing the bottom spin rather than the top one might in fact
make the phase diagram look more symmetric with respect to the
weak-coupling end, and in particular show the explicit reappearance of the
difference between regular and irregular grains in this.
On the other hand,
this would imply that we were replacing a fully frustrated column with a column
where, instead, the `reverse' field $j_n$ could yield ballistic propagation
until a finite $\xiic$ was reached -- which is not physically correct.
Thus our choice of fixing the top spin throughout the phase diagram has the
advantage of retaining the sense of the interpolation from a fully directed
model without frustration to the fully frustrated model for large $g$.

What can we predict experimentally for a box of grains?
First, the ubiquity of our four phases
(ballistic, logarithmic, activated and glassy)
throughout the phase diagram (except at the mean-field point)
vindicate our earlier picture~\cite{VI}
that the top of such a box would look ballistic,
the middle activated and the bottom glassy.
The relative sizes of the phases would of course
vary depending on frustration: we might expect, given our study, that in the
presence of strong frustration the size of the glassy phase would increase,
and that of the ballistic one decrease.
For timescales that are typical of experiment
or simulation, we would see the highest fluctuations coming from the activated
phase in the middle of a typical box~\cite{VI}; however, for much longer times
of observation, we would see large non-ergodic fluctuations at the bottom of
the box, consistent with the glassy phase.

Finally, we predict that the effects of shape are most likely to be visible
for very weak or very strong frustration, where there is a dominant
propagation of order in our column in a given direction.
In between them, when there is a balance between the propagation of order
due to gravitational settling and that due to frustration,
we might expect shape effects to disappear
as clusters of grains became the units of reorganisation.

\end{document}